\documentclass[a4paper,UKenglish]{lipics-v2016}
\usepackage{booktabs}   
\usepackage{subcaption} 
\usepackage[T1]{fontenc}
\usepackage{amsmath}
\usepackage{amssymb}
\usepackage{hyperref}
\usepackage{fancyvrb}       
\usepackage[most]{tcolorbox}
\usepackage{xcolor}
\usepackage{tikz}
\usetikzlibrary{fit} 
\usepackage{graphicx}
\usepackage{everysel}
\usepackage{float}
\usepackage{caption}
\usepackage{gensymb}
\usepackage{flushend}  

\EverySelectfont{%
\fontdimen2\font=0.4em
\fontdimen3\font=0.2em
\fontdimen4\font=0.1em
\fontdimen7\font=0.1em
}

\EventLongTitle{The 9th ACM SIGPLAN International Conference on Certified Programs and Proofs (CPP 2020)}
\EventShortTitle{CPP 2020}
\EventAcronym{CPP}
\EventYear{2020}
\EventDate{Jan 20--21, 2020}
\EventLocation{New Orleans, United States}

\begin{document}

\title{A Constructive Formalization of the Weak Perfect Graph Theorem}  

\author[1]{Abhishek Kr Singh}
\affil[1]{Tata Institute of Fundamental Research, Mumbai, India\\
	\texttt{abhishek.uor@gmail.com}}
\author[2]{Raja Natarajan}
\affil[2]{Tata Institute of Fundamental Research, Mumbai, India\\
	\texttt{raja@tifr.res.in}}
\authorrunning{A. K. Singh and N. Raja} 
      
\newcommand{\tw}{\texttt}
\newcommand{\rw}{\rightarrow}
\newcommand{\lrw}{\leftrightarrow}
\newcommand{\Rw}{\Rightarrow}

\newtheorem{Def}{Definition}
\newtheorem{Claim}{Claim}
\newtheorem{Conj}{Conjecture}
\newtheorem{Obs}{Observation}
\newtheorem{Step}{S\hspace{-0.3em}}

\definecolor{vlgray}{gray}{1}
\def \d {1}
\tikzset{
vertR/.style={scale=0.4,circle, draw=red!150, fill=red!100,thick},
vertB/.style={scale=0.4,circle, draw=blue!150, fill=blue!100,thick},
vertG/.style={scale=0.4,circle, draw=green!150, fill=green!100,thick},
vertY/.style={scale=0.4,circle, draw=yellow!250, fill=yellow!200,thick},
vertBlack/.style={scale=0.3,circle, draw=black!150, fill=black!100,thick}, 
vertGB/.style={scale=0.3,circle, draw=black!60, fill=black!40,thick}, 
vertC/.style={scale=0.4,circle,draw=cyan!150,fill=cyan!100,thick}}

\newcommand{\UTriangle}
{
\path(0,0) coordinate (Origin);
\path (Origin)+ (0:0) coordinate(A) +(0: \d) coordinate(B) + (60: \d) coordinate(C);
\draw (A)--(B)--(C)--(A);
\node (a) at (A) [vertR] {};
\node (b) at (B) [vertB] {};
\node (c) at (C) [vertG] {};
}

%
\newcommand{\USquare}
{
\path(2,0) coordinate (Origin);
\path(Origin) ++(0:0) coordinate(A) ++(0:\d) coordinate(B) ++ (90:\d) coordinate(C) ++ (180:\d) coordinate(D);
\draw (A)--(B)--(C)--(D)--(A);
\node (a) at (A)[vertR]{};
\node (b) at (B)[vertB]{};
\node (c) at (C)[vertR]{};
\node (d) at (D)[vertB]{};
}

%
%
\newcommand{\URectangle}
{
\path(0,0) coordinate (Origin);
\path(Origin) ++(0:0) coordinate(A) ++(0:\d+1) coordinate(B) ++ (90:\d) coordinate(C) ++ (180:\d+1) coordinate(D);
\draw (A)--(B)--(C)--(D)--(A);
\node (a) at (A)[vertR]{};
\node (b) at (B)[vertB]{};
\node (c) at (C)[vertR]{};
\node (d) at (D)[vertB]{};
}

\newcommand{\UPentagon}
{
\path(0,0) coordinate (Origin);
\path (Origin) ++ (0:0) coordinate(Current);
\foreach \i in{0,1,2,...,4}
{
\path (Current)++(0:0) coordinate(V\i);
\path (Current)++(\i*72: \d) coordinate(Current);
}
\draw (V0)--(V1)--(V2)--(V3)--(V4)--(V0);
\node (v0) at (V0) [vertR]{};
\node (v1) at (V1) [vertB]{};
\node (v2) at (V2) [vertR]{};
\node (v3) at (V3) [vertB]{};
\node (v4) at (V4) [vertG]{};
}

\newcommand{\UStar}
{
\path(2,0) coordinate (Origin);
\path (Origin) ++ (0:0) coordinate(A) ++ (0:\d) coordinate(B) ++ (72:\d) coordinate(C) ++ (2*72:\d) coordinate(D) ++ (3*72:\d) coordinate(E);

\draw (A)--(C);
\draw (B)--(D);
\draw (C)--(E);
\draw (B)--(E);
\draw (A)--(D);
\node (a)[vertR] at (A){};
\node (b)[vertB] at (B){};
\node (c)[vertB] at (C){};
\node (d)[vertG] at (D){};
\node (e)[vertR] at (E){};
}

\newcommand{\UHouseComp}
{
\path(3,0) coordinate (Origin);
\path (Origin) ++ (0:\d) coordinate(Current);
\foreach \i in{1,2,...,5}
{
\path (Current)++(0:0) coordinate(V\i);
\path (Current)++(\i*72: \d) coordinate(Current);
}

\draw (V4)--(V1)--(V3)--(V5)--(V2);

\node (v1) at (V1) [vertB]{};
\node (v2) at (V2) [vertR]{};
\node (v3) at (V3) [vertR]{};
\node (v4) at (V4) [vertR]{};
\node (v5) at (V5) [vertB]{};
}

\newcommand{\UDoublepentagon}
{
\path(5,0) coordinate (Origin);
\path (Origin) ++ (0:0) coordinate(Current);
\foreach \i in{0,1,2,...,4}
{
\path (Current)++(0:0) coordinate(V\i);
\path (Current)++(\i*72: \d) coordinate(Current);
}
\draw (V0)--(V1)--(V2)--(V3)--(V4)--(V0);
\path(8,0)coordinate(Origin);
\path (Origin) ++ (0:0) coordinate(Current);
\foreach \i in{0,1,2,...,4}
{
\path (Current)++(0:0) coordinate(U\i);
\path (Current)++(\i*72: \d) coordinate(Current);
}
\draw (U0)--(U1)--(U2)--(U3)--(U4)--(U0);
\draw (V2)--(U4)--(V3)--(U3)--(V2);
\node (v0) at (V0) [vertR]{};
\node (v1) at (V1) [vertB]{};
\node (v2) at (V2) [vertR]{};
\node (v3) at (V3) [vertB]{};
\node (v4) at (V4) [vertG]{};
\node (u0) at (U0) [vertC]{};
\node (u1) at (U1) [vertY]{};
\node (u2) at (U2) [vertC]{};
\node (u3) at (U3) [vertY]{};
\node (u4) at (U4) [vertBlack]{};
}


\newcommand{\UBipartitegraph}
{
\path(6,0)coordinate(Origin);
\def \dy  {\d/2}
\path(Origin) ++ (0,0) coordinate(l1) ++ (0,\dy) coordinate(l2) ++ (0,\dy) coordinate(l3) ++ (1,0) coordinate(r3) ++ (0,-\dy) coordinate(r2) ++ (0,-\dy) coordinate(r1);
\draw  (l1)--(r1);
\draw (l1)--(r2);
\draw (l2)--(r3);
\draw (l3)--(r2);

\node ()[vertR] at (l1){};
\node ()[vertR] at (l2){};
\node ()[vertR] at (l3){};

\node ()[vertB] at (r1){};
\node ()[vertB] at (r2){};
\node ()[vertB] at (r3){};
}


\newcommand{\ULine}
{
\path (0,0) coordinate(Origin);
\path(Origin) +(0,0) coordinate(v1)+ (\d,0)coordinate(v2);
\draw(v1)--(v2);
\node ()[vertR] at(v1){};
\node ()[vertB] at(v2){};
}

\newcommand{\USkewedpentagon}
{
\path(0,0) coordinate(Origin);
\path(Origin) ++ (0:0) coordinate(A) ++ (0:\d) coordinate(B) ++ (1*72:\d) coordinate(C) ++ (2*72:\d) coordinate(D) ++ (3*72:\d) coordinate(E);
\draw (A)--(C)--(D)--(E)--(B)--(A);
\node() [vertR] at (A){};
\node() [vertG] at (B){};
\node() [vertB] at (C){};
\node() [vertR] at (D){};
\node() [vertB] at (E){};
}

\newcommand{\UExtendedpentagon}
{
\path(0,0) coordinate (Origin);
\path (Origin) ++ (0:0) coordinate(Current);
\foreach \i in{0,1,2,...,4}
{
\path (Current)++(0:0) coordinate(V\i);
\path (Current)++(\i*72: \d) coordinate(Current);
}
\path (V2) ++ (25:2*\d/3) coordinate (V5);
\draw (V0)--(V1)--(V2)--(V3)--(V4)--(V0);
\draw (V1)--(V5)--(V3);
\draw (V2)-- (V5);
\node (v0) at (V0) [vertR]{};
\node (v1) at (V1) [vertB]{};
\node (v2) at (V2) [vertR]{};
\node (v3) at (V3) [vertB]{};
\node (v4) at (V4) [vertG]{};
\node (v5) at (V5) [vertG]{};
}
\newcommand{\LGraph}
{ 
\path(0,0) coordinate(Origin);
\path(Origin) ++ (0:0) coordinate(V2) ++ (5:0.7*\d) coordinate(V3) ++ (100:2*\d) coordinate(A) ++ (235:1.25*\d) coordinate(V1);
\draw (V2)--(A)--(V1);
\draw (V3)--(A);
\node (v2) at (V2) [vertBlack,label={below:{$v_{2}$}}]{};
\node (v3) at (V3) [vertBlack,label={below:{$v_{3}$}}]{};
\node (a) at (A) [vertBlack,label={above:{$a$}}]{};
\node (v1) at (V1) [vertBlack,label={below:{$v_{1}$}}]{};
\path (V2) ++ (280:1.5*\d) coordinate (G);
\node (g) at (G) []{$G$};

\path(2.2*\d,0)coordinate(Origin);
\path(Origin) ++ (0:0) coordinate(V2) ++ (5:0.7*\d) coordinate(V3) ++ (100:2*\d) coordinate(A) ++ (235:1.25*\d) coordinate(V1);
\path (V3) ++ (75:1.1*\d) coordinate (A'); 
\draw (V2)--(A)--(V1);
\draw (V3)--(A);
\draw [thin,gray!40] (V2)--(A')--(V1);
\draw [thin,gray!40] (V3)--(A')--(A);
\node (v2) at (V2) [vertBlack,label={below:{$v_{2}$}}]{};
\node (v3) at (V3) [vertBlack,label={below:{$v_{3}$}}]{};
\node (a') at (A') [vertGB, label={right:{$a^{\prime}$}}]{};
\node (a) at (A) [vertBlack,label={above:{$a$}}]{};
\node (v1) at (V1) [vertBlack,label={below:{$v_{1}$}}]{};

\path (V2) ++ (280:1.5*\d) coordinate (G');
\node (g') at (G') []{$G'$};

\path(5*\d,0)coordinate(Origin);
\path(Origin) ++ (0:0) coordinate(V2) ++ (5:0.7*\d) coordinate(V3) ++ (100:2*\d) coordinate(A) ++ (235:1.25*\d) coordinate(V1);
\path (V3) ++ (75:1.3*\d) coordinate (A'); 
\path (A') ++ (100:0.8*\d) coordinate (A'');
\draw (V2)--(A)--(V1);
\draw (V3)--(A);
\draw [thin,gray!40] (V2)--(A')--(V1);
\draw [thin,gray!40] (V3)--(A')--(A);

\draw [thin,gray!40] (V2)--(A'')--(V1);
\draw [thin,gray!40] (V3)--(A'')--(A);
\draw [thin,gray!40] (A'')--(A');

\node (v2) at (V2) [vertBlack,label={below:{$v_{2}$}}]{};
\node (v3) at (V3) [vertBlack,label={below:{$v_{3}$}}]{};
\node (a') at (A') [vertGB,label={right:{$a^{\prime}$}}]{};
\node (a'') at (A'') [vertGB, label={right:{$a^{\prime\prime}$}}]{};
\node (a) at (A) [vertBlack,label={above:{$a$}}]{};
\node (v1) at (V1) [vertBlack,label={below:{$v_{1}$}}]{};

\path (V2) ++ (280:1.5*\d) coordinate (G'');
\node (g'') at (G'') []{$G''$};
}

\newcommand{\GenGExp}{
\path(0,0)coordinate(origin);
\path(origin) ++ (0:0) coordinate(D) ++ (0:1.5*\d) coordinate(C) ++ (90:\d) coordinate(B);
\path(origin) ++ (90:1.5*\d) coordinate(A);
\draw (D)--(C)--(B)--(A)--(D);
\node (d) at (D) [vertBlack,label={below left:{$d$}}]{};
\node (c) at (C) [vertBlack,label={below right:{$c$}}]{};
\node (b) at (B) [vertBlack,label={above right:{$b$}}]{};
\node (a) at (A) [vertBlack,label={above left:{$a$}}]{};

\path(origin) ++ (320:\d) coordinate(G);
\node (g) at (G) [label={below : {$G$}}]{};

\draw [->, thick] (4.2, 0.6*\d) to [out=170, in = 20] (2.8, 0.5*\d);
\node (fg) at (3.5, 0.6*\d) [label={above:{$g$}}]{};

\path(5,0) coordinate(origin);
\path(origin) ++ (0:0) coordinate(D) ++ (0:1.5*\d) coordinate(C) ++ (90:\d) coordinate(B);
\path(origin) ++ (90:1.5*\d) coordinate(A);
\path(A) ++ (30:0.25*\d) coordinate (A2);
\path(A) ++ (210:0.25*\d) coordinate (A1);
\path(D) ++ (270:0.2) coordinate (D1);
\path(C) ++ (0,0) coordinate (C1) ++ (0:0.5*\d) coordinate (C2) ++ (270:0.5*\d) coordinate (C3) ++ (180:0.5*\d) coordinate (C4);
\path(B) ++ (0:0) coordinate (B1) ++ (20:0.5*\d) coordinate (B2) ++ (140:0.5*\d) coordinate (B3);

\foreach \i in{1,2}
{
\foreach \j in {1,2,3}
{ \draw [gray!40](A\i)--(B\j); } 
}
\foreach \i in{1,2,3}
{
\foreach \j in {1,2,3,4}
{ \draw [gray!40](B\i)--(C\j); } 
}
\foreach \i in{1}
{
\foreach \j in {1,2,3,4}
{ \draw [gray!40](D\i)--(C\j); } 
}
\foreach \i in{1,2}
{
\foreach \j in {1}
{ \draw [gray!40](A\i)--(D\j); } 
}

\node (a1) at (A1) [vertBlack,label={above : {$V_{a}$}}]{};
\node (a2) at (A2) [vertBlack]{};
\node (b1) at (B1) [vertBlack]{};
\node (b2) at (B2) [vertBlack, label={above : {$V_{b}$}}]{};
\node (b3) at (B3) [vertBlack]{};
\node (c1) at (C1) [vertBlack]{};
\node (c2) at (C2) [vertBlack]{};
\node (c3) at (C3) [vertBlack, label={right : {$V_{c}$}}]{};
\node (c4) at (C4) [vertBlack]{};
\node (d1) at (D1) [vertBlack, label={below : {$V_{d}$}}]{};
\draw (A1)--(A2);
\draw (B1)--(B2)--(B3)--(B1);
\draw (C1)--(C2)--(C3)--(C4)--(C1)--(C3);
\draw (C2)--(C4);
\path(origin) ++ (320:\d) coordinate(G');
\node (g') at (G') [label={below : {$G'$}}]{};
}

\newcommand{\LPolygonABCD}
{
\path(0,0)coordinate(origin);
\path(origin) ++ (0:0) coordinate(D) ++ (0:1.5*\d) coordinate(C) ++ (90:\d) coordinate(B);
\path(origin) ++ (90:1.5*\d) coordinate(A);
\draw (D)--(C)--(B)--(A)--(D);
\node (d) at (D) [vertBlack,label={below left:{$d$}}]{};
\node (c) at (C) [vertBlack,label={below right:{$c$}}]{};
\node (b) at (B) [vertBlack,label={above right:{$b$}}]{};
\node (a) at (A) [vertBlack,label={above left:{$a$}}]{};

\path(origin) ++ (320:\d) coordinate(G1);
\node (g1) at (G1) [label={below : {$G_{1}$}}]{};
}

\newcommand{\LPolygonVaBCD}
{
\path (2.5,0) coordinate (Origin);
\path(Origin) ++ (0:0) coordinate(D) ++ (0:1.5*\d) coordinate(C) ++ (90:\d) coordinate(B);
\path(Origin) ++ (90:1.5*\d) coordinate(A);
\path(A) ++ (30:0.25*\d) coordinate (A2);
\path(A) ++ (210:0.25*\d) coordinate (A1);
\foreach \i in{1,2}
{
\draw [gray!40](A\i)--(D); 
}
\foreach \i in{1,2}
{
\draw [gray!40](A\i)--(B);  
}
\node (a1) at (A1) [vertBlack,label={above : {$V_{a}$}}]{};
\node (a2) at (A2) [vertBlack]{};
\node (b) at (B) [vertBlack, label={above : {$b$}}]{};
\node (c) at (C) [vertBlack, label={below : {$c$}}]{};
\node (d) at (D) [vertBlack, label={below : {$d$}}]{};
\draw (A1)--(A2);
\draw (D)--(C)--(B);
\path(Origin) ++ (320:\d) coordinate(G2);
\node (g2) at (G2) [label={below : {$G_{2}$}}]{};
}

\newcommand{\LPolygonVaVbCD}
{
\path(5,0)coordinate(origin);
\path(origin) ++ (0:0) coordinate(D) ++ (0:1.5*\d) coordinate(C) ++ (90:\d) coordinate(B);
\path(origin) ++ (90:1.5*\d) coordinate(A);
\path(A) ++ (30:0.25*\d) coordinate (A2);
\path(A) ++ (210:0.25*\d) coordinate (A1);
\path(B) ++ (0:0) coordinate (B1) ++ (20:0.5*\d) coordinate (B2) ++ (140:0.5*\d) coordinate (B3);
\foreach \i in{1,2}
{
\foreach \j in {1,2,3}
{ \draw [gray!40](A\i)--(B\j); } 
}
\foreach \i in{1,2,3}
{
\draw [gray!40](B\i)--(C); 
}
\foreach \i in{1,2}
{
\draw [gray!40](A\i)--(D); 
}
\node (a1) at (A1) [vertBlack,label={above : {$V_{a}$}}]{};
\node (a2) at (A2) [vertBlack]{};
\node (b1) at (B1) [vertBlack]{};
\node (b2) at (B2) [vertBlack, label={above : {$V_{b}$}}]{};
\node (b3) at (B3) [vertBlack]{};
\node (c) at (C) [vertBlack, label={below : {$c$}}]{};
\node (d) at (D) [vertBlack, label={below : {$d$}}]{};
\draw (A1)--(A2);
\draw (B1)--(B2)--(B3)--(B1);
\draw (C)--(D);
\path(origin) ++ (320:\d) coordinate(G3);
\node (g3) at (G3) [label={below : {$G_{3}$}}]{};
}

\newcommand{\LExtendedpolygon}
{
\path(8,0)coordinate(origin);
\path(origin) ++ (0:0) coordinate(D) ++ (0:1.5*\d) coordinate(C) ++ (90:\d) coordinate(B);
\path(origin) ++ (90:1.5*\d) coordinate(A);
\path(A) ++ (30:0.25*\d) coordinate (A2);
\path(A) ++ (210:0.25*\d) coordinate (A1);
\path(D) ++ (270:0.2) coordinate (D1);
\path(C) ++ (0,0) coordinate (C1) ++ (0:0.5*\d) coordinate (C2) ++ (270:0.5*\d) coordinate (C3) ++ (180:0.5*\d) coordinate (C4);
\path(B) ++ (0:0) coordinate (B1) ++ (20:0.5*\d) coordinate (B2) ++ (140:0.5*\d) coordinate (B3);

\foreach \i in{1,2}
{
\foreach \j in {1,2,3}
{ \draw [gray!40](A\i)--(B\j); } 
}
\foreach \i in{1,2,3}
{
\foreach \j in {1,2,3,4}
{ \draw [gray!40](B\i)--(C\j); } 
}
\foreach \i in{1}
{
\foreach \j in {1,2,3,4}
{ \draw [gray!40](D\i)--(C\j); } 
}
\foreach \i in{1,2}
{
\foreach \j in {1}
{ \draw [gray!40](A\i)--(D\j); } 
}

\node (a1) at (A1) [vertBlack,label={above : {$V_{a}$}}]{};
\node (a2) at (A2) [vertBlack]{};
\node (b1) at (B1) [vertBlack]{};
\node (b2) at (B2) [vertBlack, label={above : {$V_{b}$}}]{};
\node (b3) at (B3) [vertBlack]{};
\node (c1) at (C1) [vertBlack]{};
\node (c2) at (C2) [vertBlack]{};
\node (c3) at (C3) [vertBlack, label={right : {$V_{c}$}}]{};
\node (c4) at (C4) [vertBlack]{};
\node (d1) at (D1) [vertBlack, label={below : {$V_{d}$}}]{};
\draw (A1)--(A2);
\draw (B1)--(B2)--(B3)--(B1);
\draw (C1)--(C2)--(C3)--(C4)--(C1)--(C3);
\draw (C2)--(C4);
\path(origin) ++ (320:\d) coordinate(G4);
\node (g4) at (G4) [label={below : {$G_{4}$}}]{};
}

\newcommand{\GraphConstruct}
{
\path(2,0)coordinate(origin);
\path(origin) ++ (0:0) coordinate(B) ++ (90:2*\d) coordinate(A) ++ (90:0.8*\d) coordinate(Au);
\path(A) ++ (45: 0.8*\d) coordinate(Ar);
\path(A) ++ (135: 0.8*\d) coordinate(Al);
\path(B) ++ (45: 0.8*\d) coordinate(Br);
\path(B) ++ (135: 0.8*\d) coordinate(Bl);
\path [fill= gray!10, rotate=135] (A) ellipse (25pt and 7pt);
\path [fill= gray!10, rotate=45] (A) ellipse (25pt and 7pt);
\path [fill= gray!10, rotate=90] (A) ellipse (25pt and 7pt);
\path [fill= gray!10, rotate=45] (B) ellipse (25pt and 7pt);
\path [fill= gray!10, rotate=135] (B) ellipse (25pt and 7pt);

\node (v1) at (Ar) [label={right :{$I_1$}}]{};
\node (v2) at (Au) [label={above :{$I_2$}}]{};
\node (v3) at (Al) [label={left :{$I_3$}}]{};
\node (v4) at (Br) [label={right:{$I_5$}}]{};
\node (v5) at (Bl) [label={left:{$I_4$}}]{};
\node (v6) at (B) [vertBlack,label={below:{$b$}}]{};
\node (v7) at (A) [vertBlack,label={below:{$a$}}]{};
\path(origin) ++ (270:0.7*\d) coordinate(C);
\node (c) at (C) [label={below : {$C$}}]{};

\path(6,0) coordinate (origin);
\path (origin) ++ (0:0) coordinate(Bl) ++ (0:1.5*\d) coordinate(Br)++(72:1.5*\d) coordinate(Ar);
\path (Ar) ++ (2*72:1.5*\d) coordinate(Au)++ (3*72:1.5*\d) coordinate(Al);

\path [fill= gray!10, rotate=135] (Al) ellipse (25pt and 7pt);
\path [fill= gray!10, rotate=45] (Ar) ellipse (25pt and 7pt);
\path [fill= gray!10, rotate=90] (Au) ellipse (25pt and 7pt);
\path [fill= gray!10, rotate=45] (Bl) ellipse (25pt and 7pt);
\path [fill= gray!10, rotate=135] (Br) ellipse (25pt and 7pt);

\node (v1) at (Ar) [label={above right :{$I_{1}'$}}]{};
\node (v2) at (Au) [label={above left:{$I_{2}'$}}]{};
\node (v3) at (Al) [label={above left :{$I_{3}'$}}]{};
\node (v4) at (Br) [label={right:{$I_{4}'$}}]{};
\node (v5) at (Bl) [label={left:{$I_{5}'$}}]{};

\draw [gray!40](Al)--(Ar)--(Au)--(Al); 
\draw [gray!40](Bl)--(Br);

\node (v11) at (Ar) [vertBlack,label={below:{$(a,1)$}}]{};
\node (v22) at (Au) [vertBlack,label={right:{$(a,2)$}}]{};
\node (v33) at (Al) [vertBlack,label={below:{$(a,3)$}}]{};
\node (v44) at (Bl) [vertBlack,label={below:{$(b,5)$}}]{};
\node (v55) at (Br) [vertBlack,label={below:{$(b,4)$}}]{};
 
\path(origin) ++ (320:\d) coordinate(C');
\node (c') at (C') [label={below : {$C'$}}]{};
}

\newcommand{\LDisjStable}
{
\path(0,0)coordinate(origin);
\path(origin) ++ (0:0) coordinate(V1) ++ (0:1.5*\d) coordinate(V2) ++ (90:1.5*\d) coordinate(V3);
\path(origin) ++ (90:1.5*\d) coordinate(V4);
\path [fill= gray!10, rotate=135] (V1) ellipse (23pt and 8pt);
\path [fill= gray!10, rotate=45] (V2) ellipse (23pt and 8pt);
\path [fill= gray!10, rotate=135] (V3) ellipse (23pt and 8pt);
\path [fill= gray!10, rotate=45] (V4) ellipse (23pt and 8pt);
\foreach \i in {1,2,3,4}
{
\foreach \j in {1,2,3,4}
{ \draw [gray!40](V\i)--(V\j); } 
} 
\node (v1) at (V1) [vertBlack,label={below left:{$I_1$}}]{};
\node (v2) at (V2) [vertBlack,label={below right:{$I_2$}}]{};
\node (v3) at (V3) [vertBlack,label={above right:{$I_3$}}]{};
\node (v4) at (V4) [vertBlack,label={above left:{$I_4$}}]{};
\path(origin) ++ (320:\d) coordinate(G1);
\node (g1) at (G1) [label={below : {$G_{1}$}}]{};
}

\newcommand{\LOverStable}
{
\path(4.5,0)coordinate(origin);
\path(origin) ++ (0:0) coordinate(B) ++ (90:2*\d) coordinate(A) ++ (90:0.8*\d) coordinate(Au);
\path(A) ++ (45: 0.8*\d) coordinate(Ar);
\path(A) ++ (135: 0.8*\d) coordinate(Al);
\path(B) ++ (45: 0.8*\d) coordinate(Br);
\path(B) ++ (135: 0.8*\d) coordinate(Bl);
\path [fill= gray!10, rotate=135] (A) ellipse (25pt and 7pt);
\path [fill= gray!10, rotate=45] (A) ellipse (25pt and 7pt);
\path [fill= gray!10, rotate=90] (A) ellipse (25pt and 7pt);
\path [fill= gray!10, rotate=45] (B) ellipse (25pt and 7pt);
\path [fill= gray!10, rotate=135] (B) ellipse (25pt and 7pt);

\node (v1) at (Ar) [label={right :{$I_1$}}]{};
\node (v2) at (Au) [label={above :{$I_2$}}]{};
\node (v3) at (Al) [label={left :{$I_3$}}]{};
\node (v4) at (Br) [label={right:{$I_5$}}]{};
\node (v5) at (Bl) [label={left:{$I_4$}}]{};
\node (v6) at (B) [vertBlack,label={below:{$b$}}]{};
\node (v7) at (A) [vertBlack,label={below:{$a$}}]{};
\path(origin) ++ (270:0.7*\d) coordinate(G2);
\node (g2) at (G2) [label={below : {$G_{2}$}}]{};
}

\newcommand{\LSepStable}
{
\path(8,0) coordinate (origin);
\path (origin) ++ (0:0) coordinate(Bl) ++ (0:1.5*\d) coordinate(Br)++(72:1.5*\d) coordinate(Ar);
\path (Ar) ++ (2*72:1.5*\d) coordinate(Au)++ (3*72:1.5*\d) coordinate(Al);

\path [fill= gray!10, rotate=135] (Al) ellipse (25pt and 7pt);
\path [fill= gray!10, rotate=45] (Ar) ellipse (25pt and 7pt);
\path [fill= gray!10, rotate=90] (Au) ellipse (25pt and 7pt);
\path [fill= gray!10, rotate=45] (Bl) ellipse (25pt and 7pt);
\path [fill= gray!10, rotate=135] (Br) ellipse (25pt and 7pt);

\node (v1) at (Ar) [label={above right :{$I_{1}'$}}]{};
\node (v2) at (Au) [label={above left:{$I_{2}'$}}]{};
\node (v3) at (Al) [label={above left :{$I_{3}'$}}]{};
\node (v4) at (Br) [label={right:{$I_{4}'$}}]{};
\node (v5) at (Bl) [label={left:{$I_{5}'$}}]{};

\draw [gray!40](Al)--(Ar)--(Au)--(Al); 
\draw [gray!40](Bl)--(Br);

\node (v11) at (Ar) [vertBlack,label={below:{$a_1$}}]{};
\node (v22) at (Au) [vertBlack,label={right:{$a_2$}}]{};
\node (v33) at (Al) [vertBlack,label={below:{$a_3$}}]{};
\node (v44) at (Bl) [vertBlack,label={below:{$b_1$}}]{};
\node (v55) at (Br) [vertBlack,label={below:{$b_2$}}]{};

\path(origin) ++ (320:\d) coordinate(G2);
\node (g2) at (G2) [label={below : {$G_{2}'$}}]{};
}

\newcommand{\LPentagon}
{
\path(0,0) coordinate (Origin);
\path (Origin) ++ (0:\d) coordinate(Current);
\foreach \i in{1,2,...,5}
{
\path (Current)++(0:0) coordinate(V\i);
\path (Current)++(\i*72: \d) coordinate(Current);
}
\draw (V1)--(V2)--(V3)--(V4)--(V5)--(V1);
\node (v1) at (V1) [vertB,label=below:{$v_{1}$}]{};
\node (v2) at (V2) [vertR,label=right:{$v_{2}$}]{};
\node (v3) at (V3) [vertB,label=above:{$v_{3}$}]{};
\node (v4) at (V4) [vertG,label=left:{$v_{4}$}]{};
\node (v5) at (V5) [vertR,label=below:{$v_{5}$}]{};
\path(Origin) ++ (300:\d) coordinate(G1);
\node (g1) at (G1) []{$G_{1}$};
}

\newcommand{\LPentagonWithIs}
{
\path(0,0) coordinate (Origin);
\path (Origin) ++ (0:\d) coordinate(Current);
\foreach \i in{1,2,...,5}
{
\path (Current)++(0:0) coordinate(V\i);
\path (Current)++(\i*72: \d) coordinate(Current);
}

\path (V1) ++ (72: 0.5*\d) coordinate (V12);
\path (V4) ++ (4*72: 0.5*\d) coordinate (V45);
\path [fill= gray!10, rotate=72] (V12) ellipse (23pt and 7pt);
\path [fill= gray!10, rotate=108] (V45) ellipse (23pt and 7pt);
\path [fill= gray!10, rotate=0] (V3) ellipse (17pt and 7pt);
\node (v12) at (V12) [label=right:{$I_{1}$}]{};
\node (v45) at (V45) [label=left:{$I_{3}$}]{};
\node (v3n) at (V3) [label=right:{$I_{2}$}]{};

\draw (V1)--(V2)--(V3)--(V4)--(V5)--(V1);

\node (v1) at (V1) [vertB,label=below:{$v_{1}$}]{};
\node (v2) at (V2) [vertR,label=right:{$v_{2}$}]{};
\node (v3) at (V3) [vertB,label=above:{$v_{3}$}]{};
\node (v4) at (V4) [vertG,label=left:{$v_{4}$}]{};
\node (v5) at (V5) [vertR,label=below:{$v_{5}$}]{};
\path(Origin) ++ (300:\d) coordinate(G1);
\node (g1) at (G1) []{$G$};
}

\newcommand{\LStarWithIs}
{
\path(4,0) coordinate (Origin);
\path (Origin) ++ (0:\d) coordinate(Current);
\foreach \i in{1,2,...,5}
{
\path (Current)++(0:0) coordinate(V\i);
\path (Current)++(\i*72: \d) coordinate(Current);
}

\path (V1) ++ (72: 0.5*\d) coordinate (V12);
\path (V4) ++ (4*72: 0.5*\d) coordinate (V45);
\path [fill= gray!10, rotate=72] (V12) ellipse (23pt and 7pt);
\path [fill= gray!10, rotate=108] (V45) ellipse (23pt and 7pt);
\path [fill= gray!10, rotate=0] (V3) ellipse (17pt and 7pt);
\node (v12) at (V12) [label=right:{$I_{1}$}]{};
\node (v45) at (V45) [label=left:{$I_{3}$}]{};
\node (v3n) at (V3) [label=right:{$I_{2}$}]{};

\draw (V1)--(V3)--(V5)--(V2)--(V4)--(V1);

\node (v1) at (V1) [vertB,label=below:{$v_{1}$}]{};
\node (v2) at (V2) [vertB,label=right:{$v_{2}$}]{};
\node (v3) at (V3) [vertG,label=above:{$v_{3}$}]{};
\node (v4) at (V4) [vertR,label=left:{$v_{4}$}]{};
\node (v5) at (V5) [vertR,label=below:{$v_{5}$}]{};
\path(Origin) ++ (300:\d) coordinate(G2);
\node (g2) at (G2) []{$G'$};

}

\newcommand{\LHouse}
{
\path(0,0) coordinate (Origin);
\path (Origin) ++ (0:\d) coordinate(Current);
\foreach \i in{1,2,...,5}
{
\path (Current)++(0:0) coordinate(V\i);
\path (Current)++(\i*72: \d) coordinate(Current);
}

\draw (V1)--(V2)--(V3)--(V4)--(V5)--(V1);
\draw (V2)--(V4);

\node (v1) at (V1) [vertB,label=below:{$v_{1}$}]{};
\node (v2) at (V2) [vertR,label=right:{$v_{2}$}]{};
\node (v3) at (V3) [vertB,label=above:{$v_{3}$}]{};
\node (v4) at (V4) [vertG,label=left:{$v_{4}$}]{};
\node (v5) at (V5) [vertR,label=below:{$v_{5}$}]{};
\path(Origin) ++ (300:\d) coordinate(G1);
\node (g1) at (G1) []{$G$};
}

\newcommand{\LHouseComp}
{
\path(4,0) coordinate (Origin);
\path (Origin) ++ (0:\d) coordinate(Current);
\foreach \i in{1,2,...,5}
{
\path (Current)++(0:0) coordinate(V\i);
\path (Current)++(\i*72: \d) coordinate(Current);
}

\draw (V4)--(V1)--(V3)--(V5)--(V2);

\node (v1) at (V1) [vertB,label=below:{$v_{1}$}]{};
\node (v2) at (V2) [vertR,label=right:{$v_{2}$}]{};
\node (v3) at (V3) [vertR,label=above:{$v_{3}$}]{};
\node (v4) at (V4) [vertR,label=left:{$v_{4}$}]{};
\node (v5) at (V5) [vertB,label=below:{$v_{5}$}]{};
\path(Origin) ++ (300:\d) coordinate(G1);
\node (g1) at (G1) []{$G'$};
}

\newcommand{\LExtendedpentagon}
{
\path(3,0) coordinate (Origin);
\path (Origin) ++ (0:\d) coordinate(Current);
\foreach \i in{1,2,...,5}
{
\path (Current)++(0:0) coordinate(V\i);
\path (Current)++(\i*72: \d) coordinate(Current);
}
\path(V4) ++ (340:0.3*\d) coordinate(V4');
\draw (V1)--(V2)--(V3)--(V4)--(V5)--(V1);
\draw (V3)--(V4')--(V5);
\draw (V4)--(V4');
\node (v1) at (V1) [vertB,label=below:{$v_{1}$}]{};
\node (v2) at (V2) [vertG,label=right:{$v_{2}$}]{};
\node (v3) at (V3) [vertR,label=above:{$v_{3}$}]{};
\node (v4) at (V4) [vertB,label=left:{$v_{4}$}]{};
\node (v5) at (V5) [vertR,label=below:{$v_{5}$}]{};
\node (v4') at (V4') [vertG,label=right:{$v_{4}^{\prime}$}]{};
\path(Origin) ++ (300:\d) coordinate(G2);
\node (g2) at (G2) []{$G_{2}$};
}
%
%

\newcommand{\LDextendedpentagon}
{
\path(6,0) coordinate (Origin);
\path (Origin) ++ (0:\d) coordinate(Current);
\foreach \i in{1,2,...,5}
{
\path (Current)++(0:0) coordinate(V\i);
\path (Current)++(\i*72: \d) coordinate(Current);
}
\path(V4) ++ (340:0.3*\d) coordinate(V4');
\path(V2) ++ (200:0.3*\d) coordinate(V2');
\draw (V1)--(V2)--(V3)--(V4)--(V5)--(V1);
\draw (V3)--(V4')--(V5);
\draw (V4)--(V4');
\draw (V3)--(V2')--(V1);
\draw (V2)--(V2');
\node (v1) at (V1) [vertB,label=below:{$v_{1}$}]{};
\node (v2) at (V2) [vertG,label=right:{$v_{2}$}]{};
\node (v3) at (V3) [vertR,label=above:{$v_{3}$}]{};
\node (v4) at (V4) [vertB,label=left:{$v_{4}$}]{};
\node (v5) at (V5) [vertR,label=below
:{$v_{5}$}]{};
\node (v4') at (V4') [vertG,label=right:{$\scriptstyle{v_{4}^{\prime}}$}]{};
\node (v2') at (V2') [vertC,label=left:{$\scriptstyle{v_{2}^{\prime}}$}]{};

\path(Origin) ++ (300:\d) coordinate(G3);
\node (g3) at (G3) []{$G_{3}$};
}

\newcommand{\Pentagonsquare}
{
\path(3,0) coordinate (Origin);
\path (Origin) ++ (0:0) coordinate(Current);
\foreach \i in{0,1,2,...,4}
{
\path (Current)++(0:0) coordinate(V\i);
\path (Current)++(\i*72: \d) coordinate(Current);
}
\draw (V0)--(V1)--(V2)--(V3)--(V4)--(V0);
\node (v0) at (V0) [vertR]{};
\node (v1) at (V1) [vertB]{};
\node (v2) at (V2) [vertR]{};
\node (v3) at (V3) [vertB]{};
\node (v4) at (V4) [vertG]{};
\path(5,0.3) coordinate (Origin);
\path(Origin) ++(0:0) coordinate(A) ++(0:\d) coordinate(B) ++ (90:\d) coordinate(C) ++ (180:\d) coordinate(D);
\draw (A)--(B)--(C)--(A)--(D)--(B);
\draw(C)--(D);
\draw(V2)--(D);
\node (a) at (A)[vertG,scale=1]{};
\node (b) at (B)[vertR,scale=1]{};
\node (c) at (C)[vertB,scale=1]{};
\node (d) at (D)[vertC,scale=1]{};
}


\maketitle

\begin{abstract}
 The Perfect Graph Theorems are important results in graph theory describing the relationship between clique number  $\omega(G) $ and chromatic number $\chi(G) $ of a graph $G$.  A graph $G$ is called \emph{perfect}  if  $\chi(H)=\omega(H)$ for every induced subgraph $H$ of $G$.  The Strong Perfect Graph Theorem (SPGT) states that a graph is perfect if and only if it does not contain an odd hole (or an odd anti-hole) as its induced subgraph. The Weak  Perfect Graph Theorem (WPGT) states that a graph is perfect if and only if its complement is perfect.  In this paper, we present a formal framework for working with finite simple graphs.  We model finite simple graphs in the Coq Proof Assistant by representing its vertices as a finite set over a countably infinite domain. We argue that this approach  provides a formal framework in which it is convenient to work with different types of graph constructions (or expansions) involved in the proof of the Lov\'{a}sz Replication Lemma (LRL), which is also the key result used in the proof of Weak Perfect Graph Theorem.  Finally, we use this setting to develop a constructive formalization of the Weak Perfect Graph Theorem. 
\end{abstract}

\section{Introduction}
\label{sec:intro}
\begin{figure}[h]
\centering
\begin{tikzpicture}[scale = 1]
\UTriangle;
\UHouseComp;
\UBipartitegraph;
\end{tikzpicture}
\caption{ \label{fig:F1}Some graphs where  $\chi(G)=\omega(G)$}
\end{figure}
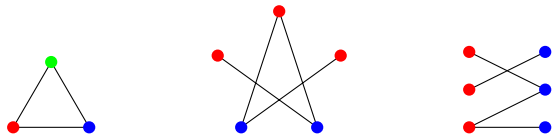

The chromatic number $\chi(G)$ of a graph $G$ is the minimum number of colours needed to colour the vertices of $G$ so that no two adjacent vertices get the same colour. Finding out the chromatic number of a graph is  NP-Hard \cite{Garey:1990:CIG:574848}.  However, an obvious lower bound for $\chi(G)$ is the clique number $\omega(G)$, the size of the biggest clique in $G$. For example, consider the  graphs shown in Figure \ref{fig:F1}.

In each of these cases the number of colours needed  is the minimum we can hope (i.e. \mbox{$\chi(G)=\omega(G)$}).  We call such graphs  \emph{nice graphs} because they admit a nice colouring (i.e. \mbox{$\chi(G)=\omega(G)$}). Can we always hope $\chi(G)=\omega(G)$ for every graph G? 

\begin{figure}[h]
\centering
\begin{tikzpicture}[scale = 1]
\UPentagon;
\UStar;
\UDoublepentagon;
\end{tikzpicture}
\caption{ \label{fig:F2} Some graphs where, $\chi(G)>\omega(G)$. For $k$ disjoint 5-cycles $\chi(G)= 3k$ and $\omega(G)= 2k$. }
\end{figure}
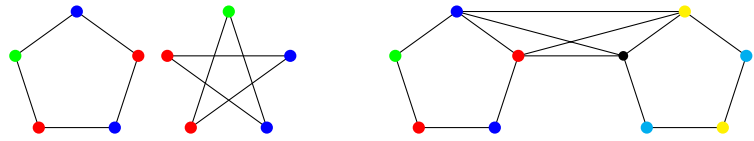

Consider the cycle of odd length 5 and its complement shown in Figure \ref{fig:F2}.  In this case one can see that  $\chi(G)=3$ and $\omega(G)=2$ (i.e. $\chi(G)>\omega(G)$).  

 In fact the gap between $\chi(G)$ and $\omega(G)$ can be made arbitrarily large. Consider the other graph shown in Figure \ref{fig:F2} which consists of two disjoint 5-cycles with all possible edges between the two cycles. This graph is a special case of a general construction where we have $k$ disjoint 5-cycles with all possible edges between any two copies. In this case one can show \cite{ChromaticGap} that $\chi(G)=3k$ but $\omega(G)=2k$.

\subsection{Perfect Graphs}
A cycle of odd length greater than or equal to 5 is called an \emph{ odd hole} and the complement of an odd hole is called an \emph{odd anti-hole}. We say that a graph $H$ is an \emph{induced subgraph} of $G$, if $H$ is a subgraph of $G$ and \hbox{$E(H)=\{uv \in E(G) \mid  u,v \in V(H)\}$}. It is interesting to note that all the graphs presented till now that do not have a nice colouring (i.e. \mbox{$\chi(G)>\omega(G)$}) have an odd hole (or odd anti-hole) as induced subgraph. 

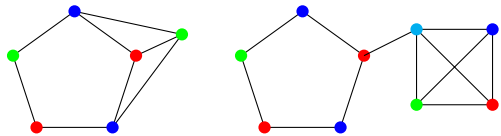
\begin{figure}[h]
\centering
\begin{tikzpicture}[scale = 1]
\UExtendedpentagon;
\Pentagonsquare;
\end{tikzpicture}
\caption{ \label{fig:F3} Graphs with $\chi(G)=\omega(G)$,  and  having odd hole as  induced subgraph.}
\end{figure}

On the other hand,  there are also some graphs containing odd holes, where $\chi(G)=\omega(G)$ (see Figure \ref{fig:F3}).  Following definition of \emph{perfect graph} avoids such constructions (similar to Figure \ref{fig:F3}), by making the idea of nice colouring hereditary.

\begin{Def}[]
A graph $G$ is called a perfect graph if \mbox{$\chi(H)=\omega(H)$} for all of its induced subgraphs $H$. 
\end{Def}

\subsection{Perfect Graph Theorems}

The perfectness of a graph $G$ can be confirmed by verifying $\chi(H)=\omega(H)$ for every induced subgraph $H$ of $G$. However, this procedure can be very costly for graphs of large order. The Perfect Graph Theorems are useful results in this direction because they provide an alternate characterisation of perfect graphs  that does not require verifying $\chi(H)=\omega(H)$ for every induced subgraph $H$.

In 1961, Claude Berge  noticed that the presence of odd holes (or odd anti-holes) as induced subgraph is the only possible obstruction for a graph to be perfect. This led him to the following conjecture known as the Strong Perfect Graph Conjecture (SPGC).

\begin{Conj}[SPGC] 
A graph is perfect if and only if it does not contain an odd hole (or an odd anti-hole) as its induced subgraph.
\end{Conj}

Berge soon realised  that this conjecture would be a hard goal to prove and  gave a weaker statement referred to as the Weak Perfect Graph Conjecture (WPGC). 
\begin{Conj}[WPGC] 
A  graph is perfect if and only if its complement is perfect.
\end{Conj}

Both  conjectures are theorems now. In 1972, Lov\'{a}sz proved a result \cite{LOVASZ1972}, known as Lov\'{a}sz Replication Lemma (LRL), which finally helped him to prove the WPGC. It took however three more decades to come up with a proof for SPGC which was finally published in a 178-page paper in 2006 by Chudnovsky et al \cite{Chudnovsky06thestrong}. 

In this paper we consider the Weak Perfect Graph Theorem and develop  a formal framework for its verification using the Coq Proof Assistant \cite{Coq:manual}.

\subsection{The Weak Perfect Graph Theorem}
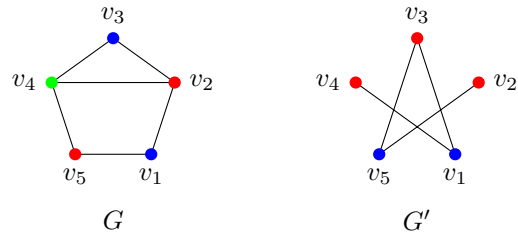
\begin{figure}[h]
\centering
\begin{tikzpicture}[scale = 1]
\LHouse;
\LHouseComp;
\end{tikzpicture}
\caption{ \label{fig:F4} $G$ is a perfect because  $G'$ is perfect. }
\end{figure}

The Weak Perfect Graph Theorem states that \emph{a graph is perfect if and only if its complement is perfect}.  For example, consider the house graph $G$ shown in Figure \ref{fig:F4}.  

It is easy to see that the complement of $G$ is an open chain which is clearly a perfect graph. Therefore, due to WPGT the house graph $G$ must be a perfect graph. 

We will now briefly look at the proof outline of Weak Perfect Graph Theorem while leaving the details for the upcoming sections.  The purpose of this discussion is to highlight the key challenges involved in the formalization of Weak Perfect Graph Theorem.  

In graph theory, a set $I$ is called a \emph{stable set} (or independent set) of graph $G$ if no two vertices of $I$ are adjacent in $G$. Following are some key insights from the proof of the Weak Perfect Graph Theorem.
 
 \begin{Obs}
 The task of proving the Weak Perfect Graph Theorem can be essentially reduced to proving that in a perfect graph $G$ there exists a clique cover  for $G$ of size $\alpha(G)$, where $\alpha(G)$ is the size of the largest stable set in $G$.
 \end{Obs}
 \begin{Obs}
For a perfect graph $G$ the existence of a clique cover of size $\alpha(G)$ can be established inductively if one can show that there exists a clique in $G$ that intersects every largest stable set of $G$.
 \end{Obs}
 
 While we leave the details and the exact lemmas establishing these observations to Section \ref{sec:WPGT},  at this point we want readers to note that the proof of  Weak Perfect Graph Theorem  reduces to proving the following claim.
 
\begin{Claim}
 In any perfect graph $G$ there exists a clique $K$ which intersects every largest stable  set of $G$.
\end{Claim}

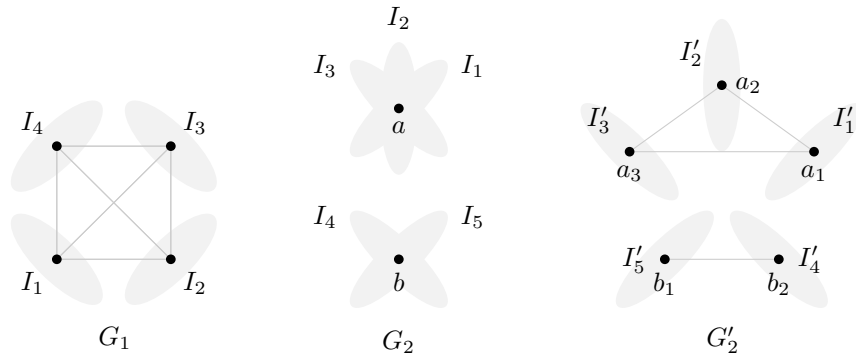
\begin{figure}[h]
\centering
\begin{tikzpicture}[scale= 1]
\LDisjStable;
\LOverStable;
\LSepStable;
\end{tikzpicture}
\caption{ \label{fig:F5} Largest independent sets (i.e. $I_1$, $I_2$, $I_3$ and $I_4$) in graph $G_1$ are mutually disjoint. $G_2$ contains some intersecting largest independent sets (e.g. independent sets $I_4$ and $I_5$ intersect at the vertex $b$).}
\end{figure}

 It is possible to prove Claim 1 by an argument employing pigeon hole principle, if every two  largest stable sets in $G$ are mutually disjoint (e.g graph $G_1$ in Figure 5). However, the same argument  does not hold if $G$ contains two or more largest independent sets with non-empty intersection (e.g. $G_2$ in Figure \ref{fig:F5}). 
 
In order to prove the claim for perfect graphs of second kind (i.e. $G_2$), Lov\'{a}sz suggested constructing a new graph $G_{2}'$ by replicating the vertices of the original graph $G_2$ in such a way that the intersecting  largest stable sets of $G_2$ become mutually disjoint in $G_{2}'$. The graph $G_{2}'$ resulting from this construction is of the first kind and hence the existence of an intersecting clique in $G_{2}'$  can be used to show a similar clique in the original graph (i.e. $G_2$). At this point, in order to complete the argument, it only remains to be proven that  the extended graph $G_{2}'$ is also a perfect graph. This can be proven using a result due to  Lov\'{a}sz  \cite{LOVASZ1972}, known as Lov\'{a}sz Replication Lemma.

\subsubsection*{Generalised Lov\'{a}sz Replication Lemma}
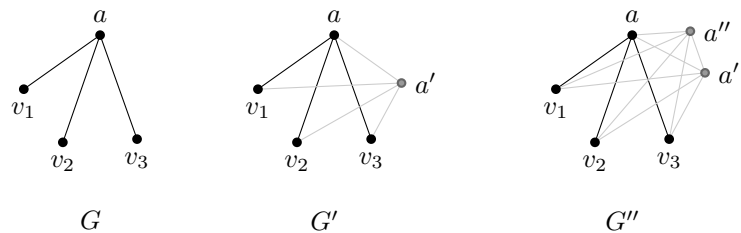
\begin{figure}[h]
\centering
\begin{tikzpicture}[xscale=1.4, yscale= 0.7]
\LGraph;
\end{tikzpicture}
\caption{ \label{fig:F6} The graph $G'$ is obtained by replicating vertex $a$ to $a'$ in $G$. Similarly the graph $G''$ is obtained by replicating vertex $a'$ to $a''$ in $G'$.}
\end{figure}

Let $G$ be a graph and  $ v \in V(G)$.  We say that $G'$ is obtained from $G$ by repeating vertex $v$ if $G'$ is obtained from $G$ by adding a new vertex $v'$ such that $v'$ is connected to $v$ and to all the neighbours of $v$ in $G$. For example, consider the graph $G'$ shown in Figure \ref{fig:F6} obtained by repeating the vertex $a$ to $a'$.

 In 1972 Lov\'{a}sz came up with the following result which states that the perfectness is preserved by replication. 
\begin{lemma}[Lov\'{a}sz 72]
 If $G'$ is obtained from a perfect graph $G$ by replicating a vertex, then $G'$ is a perfect graph.
\end{lemma} 
If we continue the process of replication we can obtain even bigger graphs where  a vertex is replaced by a clique of arbitrary size. For example, in Figure \ref{fig:F6}, the graph $G''$ is obtained from $G'$ by further repeating vertex $a'$ to a new vertex $a''$. The graph $G''$ can also be viewed as an expansion of the graph $G$ where the vertex $a$ expands into a clique of size three. Since the graph $G''$ is obtained in multiple steps from $G$ by replicating one vertex at a time we can conclude that $G''$ is perfect whenever $G$ is perfect.  

\begin{figure}[h]
\centering
\begin{tikzpicture}[scale= 0.9]
\LPolygonABCD;
\LPolygonVaBCD;
\LPolygonVaVbCD;
\LExtendedpolygon;
\end{tikzpicture}
\caption{ \label{fig:F7} The  graphs resulting from repeated replication of vertices $a$, $b$ and $c$  of $G_1$ to form cliques $V_a$ ,$V_b$ and $V_c$ of sizes 2, 3 and 4 respectively. }
\end{figure}
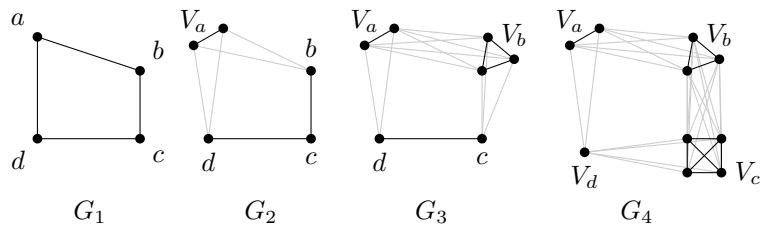

In fact, the same process of replication can be continued to obtain a graph where each vertex is replaced with a clique of arbitrary size (see Figure \ref{fig:F7}). This gives us a generalised version of the Lov\'{a}sz Replication lemma. 

\begin{lemma}[Generalised Lov\'{a}sz]
 Let $G$ be a graph and $f:V(G)\rightarrow \mathbb{N}$.  If $G'$ is a graph obtained by replacing each vertex $v_{i}$ of the graph $G$ with a complete graph of order $f(v_{i})$ then $G'$ is perfect whenever $G$ is perfect.
\end{lemma}

The generalised version of Lov\'{a}sz Replication Lemma can now be used to prove that the graph $G_{2}'$ in Figure \ref{fig:F5} is a perfect graph. The graph $G_{2}'$ is an expansion of $G_2$, where the vertices $a$ and $b$ are replaced by cliques of size 3 and 2 respectively (i.e. $f(a)=3$ and $f(b)=2$). In this case one can see that the value of function $f$ at  a vertex $v$ of $G_2$ is determined by the number of largest stable sets of $G_2$ in which $v$ appears.

\subsubsection*{Insights into the Formal Proof}
The function $f$ in the above discussions is used for describing the graph $G_{2}'$ as well as  in the statement of generalised Lov\'{a}sz Replication Lemma. Almost every mathematical presentation of the proof of Weak Perfect Graph Theorem utilises function $f$ for obtaining an intermediate graph $G'$ from $G$ and then proving the existence of an intersecting clique in $G'$. The existence of a similar clique in $G$ is then established using arguments which mostly appeal to the intuitions of the readers.

On the other hand, developing a formal proof of the Weak Perfect Graph Theorem that utilises such a function $f$ is not a fruitful endeavour. In a formal setting this approach has some serious disadvantages. For example,  it is not straightforward to formally describe the graph $G'$  based on the graph $G$ and the function $f$.  Moreover, any such description of $G'$ using  $G$ and $f$ does not provide enough information to formally prove required results about $G$ from the known results about $G'$. 

We resolve these issues by completely eliminating the use of function $f$ in the formal proof of the Weak Perfect Graph Theorem. As a result, although the overall proof idea is same, our formalization of the Weak Perfect Graph Theorem differs at many places from the usual paper proof. These differences are mostly due to the different representation of graph expansion which results in a different statement of the generalised  Lov\'{a}sz Replication lemma. 

In summary, our formalization of the Weak Perfect Graph Theorem consists of the following major steps:
\begin{Step}
Finding an alternate description for graph expansion which is free from function $f$. 
\end{Step}
\begin{Step}
Proving the generalised Lov\'{a}sz Replication lemma expressed using the new definition  of graph expansion.
\end{Step}
\begin{Step}
Finding an explicit construction to obtain the graph $G'$ so that all the largest stable sets in $G$ get separated in $G'$. Moreover, this construction should retain all the necessary information to obtain the required results about graph $G$ from the proven results about $G'$. 
\end{Step}
\begin{Step}
Proving  that the new graph $G'$ obtained from $G$ by the above construction is indeed an expansion of graph $G$ according to the definition of graph expansion.
\end{Step}

\subsection{Our Contribution}
The main contribution of this paper is the formalization of generalised Lov\'{a}sz Replication Lemma and the Weak Perfect Graph Theorem in the Coq Proof Assistant.  We also describe  the key differences between our formal proof and the usual mathematical presentation of these proofs in the literature \cite{LOVASZ1972, TrotingnonSurvey, perfectgraph}. We developed these formal proofs while extending the constructive setting of \cite{icla19} significantly. The present formalization contains many new results on the basic graph notions like complement of a graph, stable cover, clique covers, and graph expansions which were not present in \cite{icla19}.

In Section \ref{sec:FinGraphs} we describe a constructive framework to work with finite simple graphs in the Coq Proof Assistant. In this constructive setting we develop some useful results on basic graph theoretic notions like  induced subgraph, cliques, stable sets, graph isomorphism, stable cover, chromatic number, clique number and graph complements.  In Section \ref{sec:Lovasz} we provide a definition of graph expansion which is free from the function $f$. We use this definition to state and prove  the generalised Lov\'{a}sz Replication Lemma. In Section \ref{sec:WPGT}, we develop a completely constructive proof of the Weak Perfect graph theorem. We review  related work in Section \ref{sec:related}. Finally, we conclude in Section \ref{sec:conclusions} with an overview of possible future work. The Coq formalization for this paper is available at \cite{listset}.

\section{Finite Simple Graphs in Coq}
\label{sec:FinGraphs}
In our work we consider only finite simple graphs with undirected edges. Typically in a mathematical presentation the vertices of a finite graph are represented using a finite set and the edges using a binary relation on them.  In \cite{GraphDoc} Doczkal et al.  represented the vertices of finite graphs using finite types (i.e \tw{finType}) as defined in the Mathematical Components library \cite{ssreflect}.  A \emph{finite type} is a type equipped with a list enumerating all its elements. Since all the vertices come from a finite domain (i.e. \tw{finType}) almost every property on them can be represented using computable (boolean) functions. 

On the other hand in our formalization we consider the vertices of graphs  as a set whose elements belong to an instance of \tw{ordType} as described in \cite{icla19}.  An \tw{ordType} is a type equipped with a decidable equality ($=_b$) and a decidable strict total order ($<_b$). For example let \tw{A: ordType}, then following lemma ensures that the trichotomy condition holds on any two elements of \tw{A}.
\begin{lemma}
\textcolor{gray}{on\_comp} (x y: A): CompareSpec (x = y) (x <b y) (y <b x) (comp x y).
\end{lemma}

A finite graph is defined in \cite{icla19} as a dependent record with six fields.
\begin{Verbatim} 
   Record UG (A:ordType) : Type:= Build_UG {
	nodes :> list A;   nodes_IsOrd : IsOrd nodes;    
	edg: A -> A -> bool;  edg_irefl: irefl edg;   edg_sym: sym edg;  
	no_edg: edg only_at nodes   }.    	
\end{Verbatim}

The third field in the above definition is a binary relation  representing the edges of the graph. The terms \tw{(edg\_irefl G)}  and \tw{(edg\_sym G)} are proof terms whose type ensures that the edge relation of \tw{G} is irreflexive and symmetric. These restrictions on edge  make a graph \tw{G} simple and undirected.  The edge relation is considered false everywhere outside the vertices of \tw{G} which  is assured by the proof term \tw{(no\_edg G)}.  

The set of vertices  of a graph \tw{G} is represented using a list whose elements come from an infinite domain \tw{A} of type \tw{ordType}. The second field of \tw{UG} ensures that the  list  of vertices can be considered as a set. In \cite{icla19}  a finite set on \tw{A} is defined as a list whose elements are arranged using the strict order (i.e. $<_b$) of \tw{A}.

\begin{Verbatim} 
   Record set_on (A: ordType): Type := {S_of :> list A;  IsOrd_S: IsOrd S_of}. 	
\end{Verbatim}

We use this definition of graph in our formalization because it provides us a framework in which it is possible to work with different types of graph constructions in a coherent way. For example, let $G'$ be the graph obtained from a graph $G$ by replicating the vertex $a$  to $a'$ and let $G [V]$ be the induced subgraph of $G$ at the set of vertices $V$. In  mathematical discussions a vertex $v : G [V]$  is trivially considered a vertex of $G$ as well as $G'$. In this sense the vertices of all these graphs are treated in a similar way.  However, in the setting of \cite{GraphDoc}, the vertices of $G$, $G'$ and $G [V]$ all have different types.  Therefore, one needs to define generic projections between these types to convey the same information. 

The proof of the Weak Perfect Graph theorem involves many such graph constructions; both by limiting a graph to a subset of its vertices and by extending a graph while repeating some of its vertices. There are also some constructions where vertices of the resulting graph are obtained by applying various set operations on the vertices of more than one given graph. In such circumstances,  it is convenient to work with the above definition of graph where the vertices of different graphs can be finite sets from the same domain. 

 Another outcome of representing sets using ordered lists is that  we can enumerate all the subsets of a set $S$ in a list using the function $pw(S)$.  Since all the subsets of $S$ are  present in the list $pw(S)$ we can  express any predicate on power set using a boolean function on  list. This gives us a constructive framework for reasoning about properties of sets as well as their power sets.
 
  For example, consider the following definition of a boolean function \textcolor{gray}{forall\_xyb}  and its specification lemma \textcolor{gray}{forall\_xyP}.

\begin{Def}
\textcolor{gray}{forall\_xyb} (g: A$\rw$A$\rw$bool) (l: list A) := 
                 (forallb ($\lambda$ x $\Rw$ (forallb ($\lambda$ y $\Rw$ g x y) l )) l).   
\end{Def} 

\begin{lemma}
 \textcolor{gray}{forall\_xyP} (g: A$\rw$A$\rw$bool) (l: list A) :
    reflect ( $\forall$ x y, x $\in$ l $\rw$ y $\in$ l $\rw$ g x y)  (forall\_xyb g l).      
\end{lemma}
 
Note the use of \emph{reflect} predicate to specify a boolean function in the above lemma. Once a proposition $P$ is connected to a boolean $b$ using the reflect predicate it is easy to navigate between them and hence a case analysis on $P$ is possible even though the Excluded Middle principle is not provable for an arbitrary proposition in the constructive setting of Coq.

\begin{lemma}
 \textcolor{gray}{reflect\_EM} (P: Prop) (b: bool) :  reflect P b $\rw$ P $\lor$  $\sim$ P.
\end{lemma}

In fact, since the edge is a decidable relation on the set of vertices, we can  obtain decidable representations for many other important properties of graphs. For example, consider the following definitions and the corresponding lemma which gives a decidable representation to the independent sets of a graph $G$.

\begin{Def}
\textcolor{gray}{Stable} (G: UG)(I: list A):= ($\forall$ x y, x$\in$I $\rw$ y$\in$I $\rw$ edg G x y = false).  
\end{Def} 

\begin{Def}
\textcolor{gray}{stable} (G: UG)(I: list A):= (forall\_xyb ($\lambda$ x y $\Rw$ edg G x y == false) I). 
\end{Def} 

\begin{lemma}
 \textcolor{gray}{stableP} (G: UG) (I: list A): reflect (Stable G I) (stable G I).   
\end{lemma}

In a similar way we can also define subgraphs, induced subgraphs, cliques, largest clique and graph colouring as decidable predicates. Table \ref{tab:T1} shows a list of these predicates and their decidable representations.

\begin{table}[h]
\centering
\( \begin{array} {|r|l|} \hline
\tw{Propositions}   &   \tw{Boolean functions}  \\ \hline
 \tw{Subgraph G1 G2 }         &   \tw{subgraph G1 G2}  \\
 \tw{Ind\_Subgraph G1 G2 } &  \tw{ind\_subgraph G1 G2} \\
 \tw{ Stable G I }                    & \tw{stable G I }  \\
 \tw{Max\_I\_in G I}                & \tw{max\_I\_in G I}  \\
 \tw{Cliq G K }                        & \tw{cliq G K}   \\
 \tw{Max\_K\_in G K}             & \tw{ max\_K\_in G K}  \\
 \tw{Coloring\_of G f}           & \tw{coloring\_of G f} \\ \hline
\end{array} \)
\caption{\label{tab:T1} Decidable predicates on finite graphs. }
\end{table}

\subsection{Graph Colouring and Perfect Graphs}
The clique number $\omega(G)$ of a graph $G$ is  the size of biggest clique in $G$. The chromatic number $\chi(G)$ of a graph $G$ is the minimum number of colours needed to colour the vertices of $G$ so that no two adjacent vertices get the same colour. 

We have  the following lemma establishing the obvious relationship between $\chi(G)$ and $\omega(G)$.

\begin{lemma} 
\textcolor{gray}{more\_clrs\_than\_cliq\_num} (G: UG) (n: nat) (f: A$\rw$nat): cliq\_num G n $\rw$ Coloring\_of G f $\rw$ n $\leq$ |clrs\_of f G|.
\end{lemma}

Here, the expression \emph{(clrs\_of f G)} represents the set of colours used by the colouring $f$ to colour all the vertices of $G$. 
 
 We call a graph $G$ to be a \emph{nice graph} if $\chi(G)=\omega(G)$. A graph $G$ is then called a \emph{perfect graph} if every induced subgraph $H$ of $G$ is a nice graph. 

\begin{Def}
\textcolor{gray}{Nice} (G: UG) := $\forall$ n, cliq\_num G n $\rw$ chrom\_num G n.
\end{Def}
\begin{Def}
\textcolor{gray}{Perfect} (G: UG) := $\forall$ H, Ind\_subgraph H G $\rw$ Nice H.
\end{Def}

It is interesting to note that every perfect graph is also a nice graph by definition. Moreover, one can show that any induced subgraph $H$ of a perfect graph $G$ is also a perfect graph. 

\begin{lemma}\label{lem:perfect_sub_perfect}
\textcolor{gray}{perfect\_sub\_perfect}(G H: UG): Perfect G$\rw$Ind\_subgraph H G$\rw$Perfect H.
\end{lemma}

On the other hand, since $\chi(G)$ can never be smaller than $\omega(G)$, to prove that a graph is nice it is sufficient to  provide a colouring scheme $f$ which uses exactly $\omega(G)$ number of colours on $G$. 

\begin{lemma}
\textcolor{gray}{nice\_intro} (G: UG) (n:nat): cliq\_num G n $\rw$ ($\exists$ f, Coloring\_of G f $\land$ |clrs\_of f G| = n) $\rw$ Nice G.
\end{lemma}

\subsection{Graph Isomorphism}
In mathematical discussions isomorphic graphs are typically assumed to have  exactly the same properties. This idea can be made more precise using the following definition of graph isomorphism.
\begin{Def}
\textcolor{gray}{morph\_using} (f: A$\rw$B) (G: UG A) (G': UG B) :=
    (nodes G') = (img f G)  $\land$  ( $\forall$ x y, x $\in$ G $\rw$  y $\in$ G $\rw$ edg G x y = edg G' (f x) (f y)).
\end{Def}
\begin{Def}
\textcolor{gray}{iso\_using} (f: A$\rw$B) (g: B$\rw$A) (G: UG A) (G': UG B) :=
    morph\_using f G G' $\land$ morph\_using g G' G  $\land$ ( $\forall$ x, x $\in$ G $\rw$ g (f x) = x).
\end{Def}
\begin{Def}
\textcolor{gray} {iso} (G: UG A) (G': UG B) := $\exists$ (f: A$\rw$B) (g: B$\rw$A), iso\_using f g G G'.
\end{Def}
The above definition of isomorphism requires two functions (i.e. $f$ and $g$ ) to establish the isomorphism between two given graphs $G$ and $G'$. However, the following lemma states that only one function is sufficient if it is a one-to-one function on its domain.

\begin{lemma}
\textcolor{gray}{iso\_intro} (G: UG A) (G': UG B) (f: A$\rw$B):
    morph\_using f G G' $\rw$ one\_one\_on G f $\rw$ iso G G'.
\end{lemma}

Note that this representation of graph isomorphism is more general than the one defined in \cite{icla19}, which considered isomorphism between graphs whose vertices come from a single domain. Hence, we have the following additional result showing the transitivity of graph isomorphism.

\begin{lemma}
\textcolor{gray}{iso\_trans} ($G_1$: UG A) ($G_2$: UG B) ($G_3$: UG C): iso $G_1$ $G_2$ $\rw$ iso $G_2$ $G_3$ $\rw$ iso $G_1$ $G_3$.
\end{lemma}

On the other hand we have all the previous results from \cite{icla19} also available in the present setting. Probably the most important among these results is the following lemma which establishes the existence of isomorphic counterpart for an induced subgraph.

\begin{lemma}
\textcolor{gray}{iso\_subgraphs} (G H: UG A) (G': UG B) (f: A$\rw$B) (g: B$\rw$A): iso\_using f g G G' $\rw$  Ind\_subgraph H G $\rw$ ( $\exists$ H', Ind\_subgraph H' G' $\land$ iso\_using f g H H').
\end{lemma}

In a similar way we have the following lemmas showing the existence of isomorphic cliques and stable sets.

\begin{lemma}
\textcolor{gray}{iso\_cliq\_in} (G: UG A) (G': UG B) (f: A$\rw$B) (g: B$\rw$A) (K: list A):
    iso\_using f g G G' $\rw$ Cliq\_in G K $\rw$ Cliq\_in G' (img f K).
\end{lemma}

\begin{lemma}
\textcolor{gray}{iso\_stable\_in} (G: UG A) (G': UG B) (f: A$\rw$B) (g: B$\rw$A) (I: list A):
    iso\_using f g G G' $\rw$ Stable\_in G I $\rw$ Stable\_in G' (img f I).
\end{lemma}

Moreover, since same colouring can be used for isomorphic graphs we have following lemmas stating that the property of being a perfect graph is also preserved by isomorphism.

\begin{lemma}
\textcolor{gray}{chrom\_num\_G'} (G: UG A) (G': UG B) (n: nat):
    iso G G' $\rw$ chrom\_num G n $\rw$ chrom\_num G' n.
\end{lemma}

\begin{lemma}
\textcolor{gray}{nice\_G'} (G: UG A) (G': UG B) : iso G G' $\rw$ Nice G $\rw$ Nice G'.
\end{lemma}

\begin{lemma}\label{lem:isoperfect}
\textcolor{gray}{perfect\_G'} (G: UG A) (G': UG B) : iso G G' $\rw$ Perfect G $\rw$ Perfect G'.
\end{lemma}

\subsection{Stable Cover and Chromatic Number} 
A collection of stable sets (or independent sets) whose union can cover all the vertices of a graph $G$ is called a \emph{stable cover} for the graph $G$. 

\begin{Def}
\textcolor{gray} {Stable\_cover} (C: list (list A)) (G: UG A) :=
    G = (union\_over C) $\land$ ( $\forall$ I, I $\in$ C $\rw$ (Stable G I $\land$ IsOrd I)).
\end{Def}

In a similar way, a \emph{ clique cover} for $G$ is a collection of cliques whose union can cover all the vertices of $G$.

\begin{Def}
\textcolor{gray} {Cliq\_cover} (C : list (list A)) (G : UG A) :=
    G = (union\_over C) $\land$ ( $\forall$ I, I $\in$ C $\rw$ (Cliq G I  $\land$ IsOrd I)).
\end{Def}

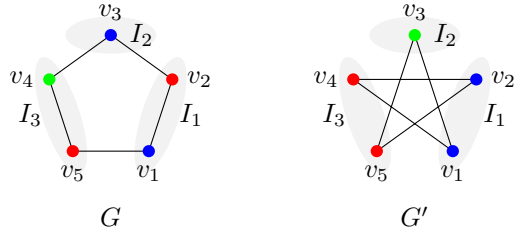
\begin{figure}[h]
\centering
\begin{tikzpicture}[scale= 1]
\LPentagonWithIs;
\LStarWithIs;
\end{tikzpicture}
\caption{\label{fig:F8} Graph $G$ and $G'$ are complement of each other. A stable cover of $G'$ is a clique cover for $G$.}
\end{figure}

 Let $G'$ be the graph shown in Figure \ref{fig:F8} and $f$ be a valid colouring for $G'$. It is easy to see that all the vertices of same colour form a stable set in $G'$. If we collect all such stable sets we get a stable cover for the whole graph $G'$.

 Therefore, we have the following result which states that from every colouring  $f$ of a graph $G$ we can obtain a stable cover $C$ of $G$ where $|C|$ is equal to the number of colours used by $f$ for colouring vertices of $G$.  

\begin{lemma}\label{lem:colortocover}
\textcolor{gray}{color\_to\_cover} (G: UG) (f: A$\rw$nat): Coloring\_of G f $\rw$
                                         ($\exists$ C, Stable\_cover C G  $\land$ |C| = |clrs\_of f G|).
\end{lemma}

It is important to note that the definition of stable cover does not require the stable sets from the collection to be mutually disjoint. Therefore, a similar argument can not be used to obtain a valid colouring $f$ from a stable set cover $C$ of $G$ which uses exactly $|C|$ colours on $G$.

 Instead we have following result which claims that there exists a valid colouring $f$ which uses at most $|C|$ colours on $G$.

\begin{lemma}
\textcolor{gray}{cover\_to\_color} (G: UG) (C: list (list A)): Stable\_cover C G $\rw$
                                                  ($\exists$ f, Coloring\_of G f  $\land$ |clrs\_of f G| $\leq$ |C|).
\end{lemma}

On the other hand  if we have a stable cover $C$ for a graph $G$ where every two stable sets in $C$ are mutually disjoint then we can obtain a colouring scheme $f$ of size $|C|$ for the graph $G$ by assigning the same colour to all the vertices of a stable set.

\begin{lemma}\label{lem:disjcovertocolor}
\textcolor{gray}{disj\_cover\_to\_color} (G: UG) (C: list (list A)):
    Stable\_cover C G $\rw$  ( $\forall$ x y, x $\in$ C $\rw$ y $\in$ C $\rw$ x $\neq$ y $\rw$ x $\cap$ y = $\phi$) $\rw$ ( $\forall$ x, x $\in$ C $\rw$ x $\neq$ $\phi$) $\rw$  ($\exists$ f, Coloring\_of G f $\land$ |clrs\_of f G| = |C|).
\end{lemma}

The above lemmas are important in the formal proof of  Weak Perfect Graph Theorem as they lead to the following result which states that a graph $G$ is nice if it can be covered using $\omega(G)$ stable sets. 

\begin{lemma}\label{lem:nice_intro1}
\textcolor{gray}{nice\_intro1} (G: UG) (n: nat): cliq\_num G n $\rw$ ($\exists$ C, Stable\_cover C G $\land$ |C| = n) $\rw$ Nice G.
\end{lemma}

It is important to note that $\omega(G)$ is also a lower bound on the size of stable covers of $G$. Assume otherwise, let $C$ be a stable cover of $G$ such that $|C| < \omega(G) $ and let $K$ be a clique in $G$ of size $\omega(G)$. Since $C$ is a stable cover for the whole graph $G$ and $|C| < |K|$, there must be a stable set in $C$ which intersects $K$ at more than one vertex. This is clearly a contradiction since no two vertices can be part of a clique and a stable set simultaneously.

\subsection{Graph Complement and the WPGT}
\label{subsec:graphcomp}
The Weak Perfect Graph Theorem is a statement about complement of a perfect graph. Let $G= (V, E)$ be a graph then the graph $G'=(V, E')$ is the complement graph of $G$ if $E' = \{(u,v): u,v \in V,   (u,v) \notin E \}$.

\begin{Def}
\textcolor{gray} {Compl} (G G': UG) :=
 (nodes G = nodes G') $\land$ ($\forall$ x y, x $\in$G $\rw$ y $\in$G $\rw$ (edg G x y $\lrw$ $\sim$edg G' x y)).
\end{Def}

For example, consider the graphs $G$ and $G'$ shown in the Figure \ref{fig:F8}. It is easy to see that these graphs are complement of each other. Also note that the stable sets in graph $G'$ (e.g $I_1$, $I_2$ and $I_3$) become cliques in the graph $G$ and vice versa. Hence, a stable cover of graph $G'$ is actually a clique cover for graph $G$. More precisely, we have the following lemmas  relating the cliques and stable sets of a graph and its complement.

\begin{lemma}
\textcolor{gray}{stable\_is\_cliq} (G G': UG A)(I: list A): Compl G G' $\rw$ Stable\_in G' I $\rw$ Cliq\_in G I.
\end{lemma}

\begin{lemma}
\textcolor{gray}{cliq\_is\_stable} (G G': UG A)(K: list A): Compl G G' $\rw$ Cliq\_in G' K $\rw$ Stable\_in G K.
\end{lemma}

\begin{lemma}\label{lem:OmegaAlfa}
\textcolor{gray}{i\_num\_cliq\_num} (G G': UG A)(n: nat): Compl G G' $\rw$ i\_num G n $\rw$ cliq\_num G' n.
\end{lemma}

At this point one can provide a justification for  \emph{Observation 1} (see  Section \ref{sec:intro}) using the results obtained so far. Let $G$ be a given perfect graph and $G'$ be its complement graph. Then, in order to prove the Weak Perfect Graph Theorem we need to prove that $G'$ is a perfect graph. 

As detailed in Section \ref{sec:WPGT}, the proof of the Weak Perfect Graph Theorem can be completed using mathematical induction on the size of $G$, where the only non trivial part of the proof is to prove that the graph $G'$ is a nice graph. According to Lemma \ref{lem:nice_intro1} this would require proving the existence of a stable cover of size $\omega(G')$ for $G'$. Again due to Lemma \ref{lem:OmegaAlfa} this is equivalent to obtaining a clique cover of size $\alpha(G)$ for the graph $G$. Therefore, proving the Weak Perfect Graph Theorem essentially reduces to proving the existence of a clique cover of size $\alpha(G)$ for a given perfect graph $G$.

\section{Generalised Lov\'{a}sz Replication Lemma}
\label{sec:Lovasz}
In this section we develop an alternate representation of graph expansion which is free from the function $f$. We use this representation of graph expansion to state and prove the generalised Lov\'{a}sz Replication Lemma. The proof of generalised Lov\'{a}sz Replication Lemma depends on definition of one vertex replication and the corresponding Lov\'{a}sz Replication Lemma.

\subsection{Vertex Replication and Lov\'{a}sz Lemma}
\begin{figure}[h]
\centering
\begin{tikzpicture}[scale = 1]
\LPentagon;
\LExtendedpentagon;
\LDextendedpentagon;
\end{tikzpicture}
\caption{ \label{fig:F9}  The graphs $G_1$, $G_2$ and $G_3$  are obtained by repeating different vertices of a cycle of length 5.}
\end{figure}
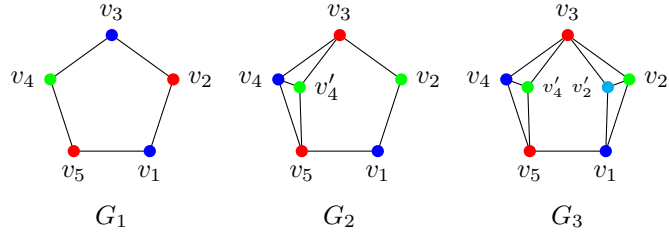

Let $G_1$, $G_2$ and $G_3$ be the graphs shown in Figure \ref{fig:F9}. The graph $G_{2}$ is obtained from $G_{1}$ by repeating vertex $v_{4}$ whereas the graph  $G_{3}$ is obtained from $G_{2}$ by repeating vertex $v_{2}$. Note that the graph $G_2$ has a nice colouring (i.e. $\chi(G_{2}) = \omega(G_{2})$), however the graph $G_3$ which is obtained by repeating vertex $v_2$ in  graph $G_2$, does not have a nice colouring (i.e. $\chi(G_{3}) > \omega(G_{3})$).  Thus, property \mbox{$\chi(G)=\omega(G)$} is not preserved by replication. 

Although the property $\chi(G)=\omega(G)$ is not preserved,  Lov\'{a}sz in 1972 came up with a surprising result which says that perfectness is preserved by replication. It states that \emph { if $G'$ is obtained from a perfect graph $G$ by replicating a vertex, then $G'$ is perfect}. Note that this result does not apply to any graph shown in Figure \ref{fig:F9}, since none of them is a perfect graph.  All of these graphs  have induced subgraph (odd hole of length 5) which does not admit a nice colouring. 

In \cite{icla19} Singh et al. developed a formalization of the Lov\'{a}sz Replication Lemma which considered repeating a single vertex of a perfect graph. We wish to generalise this approach to accommodate reasoning about repeated replication. Therefore we consider a slightly different representation for vertex replication in our formalization. 

\begin{Def}
\textcolor{gray}{Rep\_in} (G: UG) (a a': A) (G': UG) :=
    a $\in$ G $\land$ a' $\notin$ G $\land$ (nodes G') = (add a' G) $\land$ edg G' a a' $\land$
    ( $\forall$ x y, x $\in$ G $\rw$ $y \in G$ $\rw$ edg G x y = edg G' x y) $\land$ ( $\forall$ x, x $\neq$ a $\rw$ edg G x a = edg G' x a').
\end{Def}

\begin{Def}
\textcolor{gray}{Rep} G G' := $\exists$ a a', Rep\_in G a a' G'.
\end{Def}

Let the graph $G'$ be obtained by replicating vertex $a$ of $G$ to $a'$ and  let $G'-a$ represents the restriction of graph $G'$ to the set of vertices $G' \setminus \{a\}$.  Then, it is interesting to note that both the graphs $G$ and $G' - a$ are induced subgraphs of the graph $G'$. Moreover, consider a function $f$ which maps $a$ to $a'$ and $a'$ to $a$ and every other element to itself. Then the following lemma states that we can establish an isomorphism between the graphs $G$ and $G' - a$ using the function $f$.  

\begin{lemma}
\textcolor{gray}{G\_iso\_G'\_a}: iso\_usg f G ($G' - a$).
\end{lemma}

Finally, we have the following lemma stating that perfectness is preserved by replication of one vertex. 

\begin{lemma}\label{lem:RepLemma}
\textcolor{gray}{ReplicationLemma}: Perfect G $\rw$ Rep G G' $\rw$ Perfect G'.
\end{lemma}

We leave the details of formal proof of this lemma as it closely follows the proof outline presented in \cite{icla19}. However, it is important to note that the proof of this lemma involves analysis of many different cases. These cases correspond to  various predicates on sets and finite graphs. Since we have  a decidable representation for each of these predicates, we can do case analysis on them without assuming any axiom. 

\subsection{Graph Expansion and the Generalised LRL}
\label{subsec:expansion}
We need a  definition of graph expansion which can be used to state the generalised Lov\'{a}sz Replication Lemma. Moreover, we want this definition to be free from the function $f$ due to the reasons discussed in Section \ref{sec:intro}.  We consider the following definition of graph expansion which uses a backward function $g$ to remember all the essential information of the expansion. 

\begin{Def}
\textcolor{gray}{Exp\_of} (G: UG A) (G': UG B) (g: B$\rw$A) :=
    (nodes G = img g G') $\land$ ( $\forall$ x y,  x $\in$ G' $\rw$ y $\in$ G' $\rw$ x $\neq$ y $\rw$ g x = g y $\rw$ edg G' x y) $\land$ ( $\forall$ x y, x $\in$ G' $\rw$ y $\in$ G' $\rw$ g x $\neq$ g y $\rw$ edg G (g x) (g y) = edg G' x y).
\end{Def}

\begin{figure}[h]
\centering
\begin{tikzpicture}[scale = 0.9]
\GenGExp;
\end{tikzpicture}
\caption{ \label{fig:F10} Function $g$ maps vertices of cliques $V_a$, $V_b$, $V_c$ and $V_d$ to vertices $a$, $b$, $c$ and $d$ respectively.}
\end{figure}
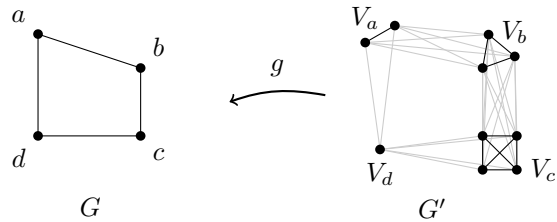

Intuitively, the function $g$ in the above definition remembers for each vertex of $G'$ its  origin in $G$. For example, see the graphs $G$ and $G'$ shown in Figure \ref{fig:F10}.

 It is important to note that the function $g$ can be considered as a homomorphism from $G'$ to $G$ which preserves the edges between vertices except when the vertices belong to a clique resulting from the expansion.   
 
 This alternate definition of graph expansion makes it easy to state and prove the following generalised Lov\'{a}sz Replication Lemma.
 
\begin{lemma}\label{lem:LovaszExpLemma}
\textcolor{gray}{LovaszExpLemma} (G: UG A) (G': UG B) (g: B$\rw$A): Exp\_of G G' g $\rw$ Perfect G $\rw$ Perfect G'.
\end{lemma}
\noindent \emph{Proof.} Using induction on the order of graph $G'$. The proof can be split into two cases. The first case corresponds to the situation where no two largest stable sets in $G$ intersect with each other. In this case $g$ is a one-to-one function on $G'$. In fact, we can show that $g$ establishes an isomorphism between $G$ and $G'$. Therefore, $G'$ is perfect since it is isomorphic to a perfect graph (Lemma \ref{lem:isoperfect}). 

The second case corresponds to the condition when there is at least one vertex, say $a$, common in two or more largest independent sets. More precisely, there exists at least two distinct vertices $x$ and $y$ in $G'$ such that $g(x) = g(y)= a$. Now, let $H'$ be the induced subgraph of $G'$ at the set of vertices $G' \setminus \{y\}$. It is easy to see that the graph $G'$ can be obtained by replicating the vertex $x$ in $H'$ to the new vertex $y$ (i.e. \emph{Rep H' G'}). Moreover, we can also show that the graph $H'$ is still an expansion of graph $G$ using the function $g$ (i.e. \emph{Exp\_of G H' g}). Hence, using induction hypothesis we can prove that $H'$ is a perfect graph. Furthermore, using Lemma \ref{lem:RepLemma} we can conclude that the graph $G'$ is a perfect graph. $\square$

\section{Formal Proof of WPGT}
\label{sec:WPGT}
We presented the overall proof outline of the Weak Perfect Graph Theorem in Section \ref{sec:intro} while leaving the exact details. In this section we develop the formal proof of Weak Perfect Graph Theorem in a bottom up way while first establishing all the essential results needed for the main lemma.

\subsection*{Existence of an Intersecting Clique}
As noted in the introduction (Section \ref{sec:intro}), the proof of \emph{Claim 1} constitutes the core segment in the formal proof of the Weak Perfect Graph Theorem. Therefore, we will first prove this result and then use it in the proof of Weak Perfect Graph Theorem. 

\begin{lemma}
In any perfect graph $G$ there exists a clique $K$ which intersects every largest stable  set of $G$.
\end{lemma}

\noindent \emph{Proof Idea}: It is not easy to see the existence of such intersecting clique in the original graph (i.e. $G$), therefore the proof of this lemma involves constructing two more graphs $G_s$ and $G_s'$. The graph $G_s$ is an induced subgraph of $G$ on those vertices which appear in at least one of the largest independent sets of $G$. Note that $G_s$ is a perfect graph since it is an induced subgraph of $G$ (Lemma \ref{lem:perfect_sub_perfect}). The aim of constructing $G_s$ is to obtain a clique $K$ in $G_s$ such that $K$ intersects every largest independent set in $G_s$. Furthermore, it can be shown that the collection of largest independent sets in $G$ is exactly the same as the collection of largest independent sets in $G_s$. Therefore, obtaining such a clique $K$ in the graph $G_s$ would automatically prove the existence of a clique that intersects every largest independent set of $G$.

It turns out that obtaining such a clique is possible by a clever use of pigeon hole principle if all the largest independent sets in $G_s$ are mutually disjoint. However, we can't assume that in the perfect graph $G_s$ all the largest independent sets are mutually disjoint. At this point, to overcome this difficulty we need another construction. We describe a construction where the intersecting largest independent sets in $G_s$ get separated in the resulting graph $G_s'$. Furthermore, we prove that the graphs $G_s$ and $G_s'$ are related by the definition of graph expansion as described in Section \ref{subsec:expansion} using a suitable function $g$.  Note that $G_s'$ is also a perfect graph since it is an expansion of the graph $G_s$ (Lemma \ref{lem:LovaszExpLemma}). Moreover, we can show that every largest independent set of $G_s$ has a corresponding largest independent set in $G_s'$. Hence an intersecting clique $K'$ in $G_s'$ can be mapped (using $g$) to an intersecting clique $K$ in $G_s$ which in turn proves the main result. $\square$

\subsubsection*{Contructing Graphs $G_s$ and $G_s'$}
We will now describe in some detail the construction of graphs $G_s$ and $G_s'$. These details are essential in understanding the formal proof of lemmas leading to the Weak Perfect Graph Theorem. Let $C$ be the collection of all the largest independent sets in $G$ and let $N$ be the union of all the sets present in $C$. Then the graph $G_s$ is the induced subgraph of $G$ at the set of vertices $N$. For instance, in our formal framework this can be achieved using the following definitions.

\begin{Verbatim} 
   Let C:= max_subs_of G (fun I => stable G I).
   Let N:= union_over C.
   Let Gs:= (ind_at N G).
\end{Verbatim}

Note that although we are calling $C$ a collection it is actually a list where each element $I \in C$ has an index which can be accessed using the expression (\tw{idx I C}). This piece of extra information is very useful in separating the intersecting independent sets in $C$. 

Let $I$ be a largest independent set in $G$ and $i$ be its index in $C$. Then we can define another set $I'$ having elements $(v,i)$ where $v \in I$. If we repeat this process for every element $I_i \in C$ we get another collection $C'$ whose elements are $I_i'$. Let $N'$ be the union of all the sets in $C'$. Now the set $N'$ can be used to represent all the vertices of the graph $G_s'$. Also note that the function $g$ here is the projection function which projects the first element of an ordered pair. In the present framework all of this can be achieved using the following definitions. 

\begin{Verbatim} 
   Let C':= mk_disj C.
   Let N':= union_over C'.
   Let g:= fun (x: A*nat) => fst x.
\end{Verbatim}

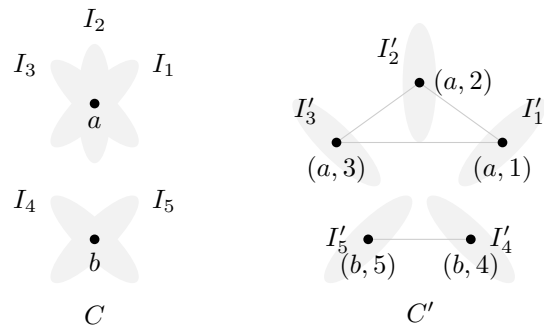
\begin{figure}[h]
\centering
\begin{tikzpicture}[scale = 0.9]
\GraphConstruct;
\end{tikzpicture}
\caption{ \label{fig:F11} Sets in $C$ gets separated in $C'$.}
\end{figure}

Here the function \tw{mk\_disj} can be seen as separating the possibly intersecting sets in $C$ to create a collection $C'$ of mutually disjoint sets. For example, let $C$ be the collection of all the largest independent sets of graph $G$ as shown in Figure \ref{fig:F11}. In this case while the vertex $a$ belongs to three largest independent sets (i.e. $I_1$, $I_2$ and $I_3$) the vertex $b$ belongs to two largest independent sets (i.e. $I_4$ and $I_5$).  

 The function \tw{mk\_disj} creates the collection $C'$ by obtaining the $I_i'$ corresponding to each $I_i \in C$. Also note that the vertices $(a,1)$, $(a,2)$ and $(a,3)$ of graph $G_s'$ can be mapped to their originating vertex $a$ in $G$ using the function $g$.
 
 In order to complete the description of graph $G_s'$ it only remains to define the edges of $G_s'$ on the set of vertices $N'$. We use the following definitions to complete the formal description of $G_s'$.

\begin{Verbatim} 
Let E1:= fun (x y: A*nat) =>
	match (g x == g y) with 
	| true => match (snd x == snd y) with
		|true => false
		|false => true
		end
	|false => E (g x) (g y)
	end.
Let E':= E1 at_ N'.               
Definition Gs' := ({| 
	nodes:= N';  
	nodes_IsOrd:= N'IsOrd;
	edg:= E';
	edg_irefl:= E'_irefl; 
	edg_sym:= E'_sym; 
	out_edg:= E'_out |}).               
\end{Verbatim}
 Now that we have a complete description of $G_s$ and $G_s'$ we can prove the following results relating the edges and vertices of these graphs.
 
 \begin{lemma}
\textcolor{gray}{C\_and\_C'}: (nodes $G_s$) = img g $G_s'$.
\end{lemma}

\begin{lemma}
\textcolor{gray}{E'\_P1}: $\forall$ x y, x $\in$ $G_s'$ $\rw$ y $\in$ $G_s'$ $\rw$ x $\neq$ y $\rw$ g x = g y $\rw$ edg $G_s'$ x y.
\end{lemma}

\begin{lemma}
\textcolor{gray}{E'\_P2}:  $\forall$ x y, x $\in$ $G_s'$ $\rw$ y $\in$ $G_s'$ $\rw$ g x $\neq$ g y $\rw$ edg $G_s$ (g x) (g y) = edg $G_s'$ x y.
\end{lemma}

With the help of these lemmas and the definition of graph expansion presented in Section \ref{subsec:expansion}, it is easy to see that the graph $G_s'$ is indeed an expansion of graph $G_s$ expressed using the function $g$.

\begin{lemma}
\textcolor{gray}{H'\_exp\_of\_H}: Exp\_of $G_s$ $G_s'$ g.
\end{lemma}

It is important to note that by now we have already described all the major steps (i.e. $S1$, $S2$, $S3$ and $S4$) of the formal proof of WPGT which were presented in the introduction (Section \ref{sec:intro}). However, it remains to show that the construction of graph $G_s'$ retains enough information to find out an intersecting clique $K'$ in $G_s'$. 

We have following lemmas establishing that the collection $C'$ is a disjoint stable cover for the graph $G_s'$.

\begin{lemma}
\textcolor{gray}{C'\_is\_disj}: $\forall$ $I_1$ $I_2$, $I_1 \in C'$ $\rw$ $I_2 \in C'$ $\rw$ $I_1$ $\neq$ $I_2$ $\rw$ $I_1 \cap I_2 = \phi$.
\end{lemma}

\begin{lemma}\label{lem:C'mem}
\textcolor{gray}{C'\_mem\_stable}: $\forall$$I'$, $I' \in C'$$\rw$Stable\_in $G_s'$ $I'$.
\end{lemma}

\begin{lemma}
\textcolor{gray}{Stable\_cover\_C'\_H'}: Stable\_cover $C'$ $G_s'$.
\end{lemma}

Additionally, we can  prove that the function $g$ is a one-to-one function on every stable set $I'$ of $G_s'$ and it maps these stable sets into stable sets of $G_s$.

\begin{lemma}
\textcolor{gray}{g\_one\_one\_on\_I'}: $\forall$ $I'$, Stable\_in $G_s'$ $I'$ $\rw$ one\_one\_on $I'$ g.
\end{lemma}

\begin{lemma}
\textcolor{gray}{img\_of\_I'\_is\_I}: $\forall$ $I'$, Stable\_in $G_s'$ $I'$ $\rw$ Stable\_in $G_s$ (img g $I'$).
\end{lemma}

In fact, the previous two Lemmas can be used together with Lemma \ref{lem:C'mem} to prove that every set in the collection $C'$ is actually a largest stable set in $G_s'$.

\begin{lemma}
\textcolor{gray}{Stable\_in\_H'}: $\forall$$I'$, $I' \in C'$ $\rw$ Max\_I\_in $G_s'$ $I'$.
\end{lemma} 

At this point we know that $C'$ is a disjoint stable cover for $G_s'$ where each entry in $C'$ is also a largest stable set in $G_s'$. Moreover, $G_s'$ is a perfect graph since it is an expansion of the perfect graph $G_s$ (Lemma \ref{lem:LovaszExpLemma}). These informations when put together can help us prove the following lemma which claims the existence of a clique $K'$ in $G_s'$ that intersects all the largest independent sets of $G_s'$.

\begin{lemma}\label{lem:K'meetsall}
\textcolor{gray}{K'\_meets\_all\_in\_C'}:$\exists K'$,Cliq\_in $G_s'$$K'$$\land$($\forall I'$, $I' \in C'$$\rw$meets $K' I'$).
\end{lemma}
\noindent \emph{Proof}. Note that all the independent sets present in $C'$ are the largest independent sets in $G_s'$. Moreover, we know that the independent sets in $C'$ form a disjoint stable cover for the whole graph $G_s'$. Therefore, we can conclude that $C'$ is the smallest stable cover possible for the graph $G_s'$. We claim that $\chi(G_s') = |C'|$. Clearly there exists a colouring $f$ for $G_s'$ which uses exactly $|C'|$ colours (Lemma \ref{lem:disjcovertocolor}). Moreover, we can show that there does not exist a colouring scheme which uses less that $|C'|$ colours. Assume otherwise and let $f'$ be a colouring which uses less than $|C'|$ colours. Then due to Lemma \ref{lem:colortocover} there exists a stable cover $C$ for $G_s'$ such that $|C| < |C'|$. This is clearly a contradiction and hence we have $\chi(G_s') = |C'|$. 

Since $G_s'$ is a perfect graph we have $\omega(G_s') = \chi(G_s')$. In other words, there exists a clique $K'$ of size $|C'|$ in the graph $G_s'$. Also note that every vertex of $K'$ appears in some $I' \in C'$. Moreover, no two vertices of $K'$ can appear in a single $I' \in C'$. This is possible only when $K'$ meets all the largest independent sets in $C'$, which completes the proof. $\square$ 

Let $K'$ be the intersecting clique assured by Lemma \ref{lem:K'meetsall}. Then one can use function $g$ and map $K'$ to a clique $K$ in $G_s$ which intersects every largest stable sets in $G_s$. Note that such a clique $K$ is also a clique in $G$ since $G_s$ is an induced subgraph of $G$.

\begin{lemma}\label{lem:Kmeetsall}
\textcolor{gray}{K\_meets\_all\_in\_C} : $\exists$ K, Cliq\_in $G$ K $\land$ ($\forall$ I, $I \in C$ $\rw$ meets K I).
\end{lemma}

\subsection*{A Clique Cover of Size $\alpha(G)$}

 The proof of Weak Perfect Graph Theorem, as remarked at the end of Section \ref{subsec:graphcomp}, essentially reduces to proving the existence of a clique cover of size $\alpha(G)$ for a given perfect graph $G$. With the help of Lemma \ref{lem:Kmeetsall} we can now prove this statement using an inductive argument.
 
\begin{lemma}\label{lem:CliqCoverExists}
\textcolor{gray}{i\_num\_cliq\_cover} (G: UG A)(n: nat):
    Perfect G $\rw$ i\_num G n $\rw$ ($\exists$ C, Cliq\_cover C G $\land$ |C| = n).
\end{lemma}
\noindent \emph{Proof}. Using induction on $n$. Let $K$ be the clique guaranteed by Lemma \ref{lem:Kmeetsall}. Let $H$ be the induced subgraph of $G$ at the set of vertices $G \setminus K$. The graph $H$ is a perfect graph since it is an induced subgraph of $G$. 

We now claim that $\alpha(H) = \alpha(G) -1$. This is true since removing $K$ decreases the size of all largest independent sets of $G$ by one in $H$. Moreover, there can't be any other independent set of size $\alpha(G)$ present in the graph $H$ since such an independent set would be a largest independent set of $G$ as well and hence will have a non empty intersection with the clique $K$. This clearly is a contradiction since $H$ and $K$ are mutually disjoint. 

Now, by the induction hypothesis we can assume that there is a clique cover $C_H$ of size $\alpha(G)-1$ for the induced subgraph $H$. Consider the collection $C = C_H \cup \{K\}$. It is easy to see that $C$ is a clique cover of size $\alpha(G)$ for the original graph $G$. $\square$

Finally, with the help of all the results developed so far we can prove our main result.

\begin{theorem}
\textcolor{gray}{wpgt} (G $G'$:  UG A): Perfect G $\rw$ Compl G $G'$ $\rw$ Perfect $G'$.
\end{theorem}
\noindent \emph{Proof}. Using induction on the order of $G$. Our goal is to prove $\omega(H')=\chi(H')$ for every induced subgraph $H'$ of $G'$. We first consider the case when $H' \neq G'$. In this case $H'$ is the complement of an induced subgraph $H$ of $G$ where $|H| < |G|$. Therefore using induction hypothesis we can conclude that $H'$ is a perfect graph since $H$ is a perfect graph. Hence, we have $\omega(H') = \chi(H')$.   

On the other hand when $H' = G'$, we have to prove $\omega(G') = \chi(G')$. In other words we have to prove that $G'$ is a nice graph. We can prove this by showing a stable cover $C'$ of size $\omega(G')$ for the graph $G'$ (Lemma \ref{lem:nice_intro1}). This in turn is equivalent to obtaining a clique cover $C$ of size $\alpha(G)$ for the graph $G$. Lemma \ref{lem:CliqCoverExists} guarantees such a clique. $\square$

\section{Related Work}
\label{sec:related}
There is no prior work known to us which formalizes any of the  Perfect graph theorems in a theorem prover. In fact there are very few formalizations of graphs in the Coq Proof Assistant. The most extensive among these is due to the formalization of four colour theorem \cite{FourColor, ssreflect} which considers only planar graphs. The work by Dufourd and Bertot on formalizing plane Delaunay triangulation \cite{TriangulationBertot}  utilises a similar notion of graphs based on hypermaps. In a recent work \cite{GraphDoc} aimed towards formalizing the graph minor theorem for treewidth two  Doczkal et al. developed a general library for graphs using the Coq Proof Assistant. This library contains many useful notions from graph theory like paths,  subgraphs, graph isomorphisms, graph minors, trees, and tree decompositions. There are also some works which formalize graphs in other theorem provers. For example, in \cite{FlySpeck} Nipkow et al.  present a formalization of planar graphs developed in Isabelle/HOL as part of the Flyspeck project. A recent work by Noschinski formalizes both simple and multi-graphs in Isabelle/HOL \cite{Noschinski2015}. Chou also developed a formalization of undirected graph theory in HOL \cite{ChouGraph}. 

We used a definition for finite graphs which is closest to the one used by Doczkal et al. \cite{GraphDoc}. However, there is an important difference in the way we define vertices of finite graphs in our formalization. We represented the vertices of finite graphs as sets over \tw{ordType} instead of \tw{finType}.  This allowed us to treat the vertices of graphs as finite sets over a countably infinite domain and  provided a framework in which it is convenient to work with different types of graph constructions in a coherent way. This difference led to an independent development of our graph library which focuses on the graph theoretic ideas relevant for the Perfect Graph Theorems.   We keep our formalization constructive by following a proof style similar to the  small scale reflections \cite{ssreflect} of ssreflect.  Moreover, since the Perfect Graph theorems are results about graph colourings and cliques, our formalization considers a different set of notions from graph theory like graph colouring, cliques, stable sets, induced subgraphs, graph complement,  chromatic number, clique number, stable set covers, and graph expansions.

\section{Conclusions and Future Works}
\label{sec:conclusions}
 We presented a fully constructive proof of the generalised Lov\'{a}sz Replication Lemma and the Weak Perfect Graph Theorem in the Coq Proof Assistant. In this process we developed a formal framework to work with finite graphs. The vertices of finite graphs are modelled as a finite set over a countably infinite domain. To keep the formalization constructive we followed a proof style similar to the  small scale reflections technique of the ssreflect \cite{ssreflect}. We use small boolean functions in a systematic way to represent various predicates over sets and graphs. These functions together with their specification lemmas helped us in avoiding  the use of Excluded-Middle in the proof development. 
 
 The same framework can also be useful in the verification of other important results on finite graphs.  An interesting direction of future work would be the formalization of the Strong Perfect Graph Theorem. This will require defining all the basic classes of graphs (e.g. \emph{bipartite graph}, \emph{line graphs} of bipartite graphs, and \emph{double split} graphs) and the decompositions (e.g. \emph{2-join}, \emph{M-join}, and the \emph{balanced skew partition}) involved in the proof of Strong Perfect Graph Theorem. A separate analysis of each of these basic notions and their related results will be very useful in the formal verification of the Strong Perfect Graph Theorem. This will not only increase our confidence in the Strong Perfect Graph Theorem but may also lead to a deeper understanding of the verified results.
 
\subsection*{Acknowledgements }
We are thankful to Kshitij Gajjar for discussing these proofs and motivating us to formalize them. We acknowledge support of the Department of Atomic Energy, Government of India, under project no. 12-R$\&$D-TFR-5.01-0500.

\bibliography{wpgt}

\end{document}